\documentclass[letterpaper]{article} 
\usepackage{aaai2027}  
\usepackage[hyphens]{url}  
\usepackage{graphicx} 
\usepackage{natbib}  
\usepackage{caption} 
\usepackage{algorithm}
\usepackage{algorithmic}
\usepackage{amsmath}
\usepackage{amssymb}
\usepackage{newfloat}
\usepackage{listings}
\DeclareCaptionStyle{ruled}{labelfont=normalfont,labelsep=colon,strut=off} 
\floatstyle{ruled}
\newfloat{listing}{tb}{lst}{}
\floatname{listing}{Listing}

\usepackage{booktabs}

\title{TI-StegoAlign: Channel-Guided Post-Training for Generative\\ Text Steganography under Tokenization Inconsistency}
\author{
    Jiuan Zhou\textsuperscript{\rm 1,\rm 2},
    Yuhao Xue\textsuperscript{\rm 1},
    Yu Cheng\textsuperscript{\rm 1,\rm 2},
    Yuan Xie\textsuperscript{\rm 1,\rm 2},
    Zhaoxia Yin\textsuperscript{\rm 1}\corresponding
}

\affiliations{
    \textsuperscript{\rm 1}East China Normal University, Shanghai, China\\
    \textsuperscript{\rm 2}Shanghai Innovation Institute, Shanghai, China\\
}

\begin{document}

\nocopyright
\maketitle

\begin{abstract}
Generative text steganography enables LLM agents to exchange secret information through task-relevant messages. Yet most methods evaluate recovery on sender-side tokens, whereas the receiver observes only surface text.  Detokenization and receiver-side retokenization can alter token boundaries, desynchronize coding states, and cause such evaluation to overestimate receiver-side recovery. Existing remedies rely on inference-time filtering or verification, correcting individual outputs without adapting the generation policy to the receiver-side channel. To address these limitations, we propose TI-StegoAlign, a channel-guided post-training framework. The
Bit-Consistent Supervised Objective (BCSO) enlarges local coding margins at realized sender-side embedding positions. Channel-Conditioned Preference Optimization (CCPO) then aligns complete stegotexts using receiver-realistic recovery, text quality, and anti-steganalysis feedback. TI-StegoAlign updates only LoRA parameters and requires no tokenization-specific correction during communication. Experimental results show 100\% receiver bit accuracy. Compared with the strongest baselines, TI-StegoAlign achieves a 21.6\% reduction in normalized perplexity deviation and a 6.3\% relative improvement in anti-steganalysis performance.

\end{abstract}

\section{Introduction}

Large language models increasingly underpin agents that communicate through dialogue \cite{wu2024autogen}, decompose complex tasks and delegate subtasks to specialized agents \cite{hong2024metagpt}, and coordinate within multi-agent teams \cite{chen2024agentverse}. These interactions are commonly mediated by natural-language messages that may be accessible to platform operators, automated monitoring systems, or other third parties \cite{motwani2024secret,huang2026whispering}. When semantically coherent messages carry recoverable information while preserving their overt conversational function, they can establish a covert channel within the interaction. Such communication may conceal the presence or content of sensitive exchanges from observers, but it can also facilitate unauthorized information transfer beyond established oversight mechanisms. Recent studies have investigated steganographic collusion \cite{motwani2024secret}, its emergence under misspecified training incentives \cite{mathew2025hidden}, metaphor-driven covert communication \cite{xu2025comet}, and event-driven covert communication protocols \cite{huang2026whispering}.

\begin{figure}[t]
    \centering
    \includegraphics[width=1.0\linewidth]{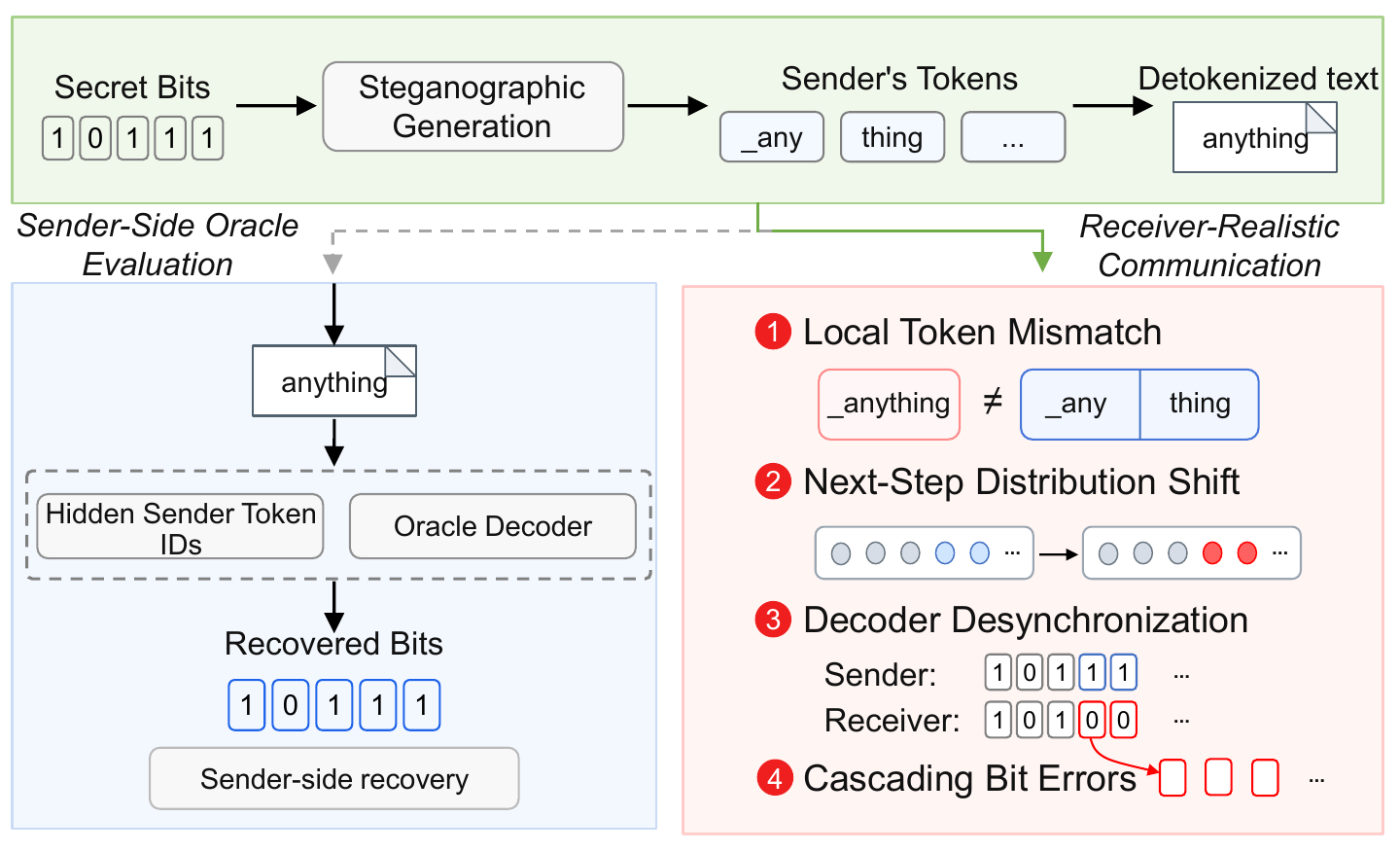}
    \caption{Sender-side oracle evaluation reuses the hidden generation tokens, whereas receiver-realistic communication extracts from retokenized surface text. A local boundary change can desynchronize subsequent coding states and cause cascading bit errors.}
    \label{fig:pipeline}
\end{figure}

Generative text steganography embeds secret information during language generation while preserving the task-relevant semantics of agent messages. Early work combined neural language models with token-level coding \cite{yang2019rnnstega,ziegler2019neural,shen2020near,zhang2021provably}, followed by advances in distributional security \cite{kaptchuk2021meteor,schroederdewitt2023perfectly,ding2023discop}, sampling efficiency \cite{wang2025sparsamp}, and public-key constructions \cite{zhang2025publickey}; recent LLM-based methods have explored black-box embedding \cite{wu2024generative}, adaptive strategy evolution \cite{zhou2026auto}, and distribution optimization \cite{huang2026odstega}. In token-sampling schemes, correct extraction requires the sender and receiver to operate on the same token sequence. Yet receiver-side retokenization may produce a different sequence because of segmentation ambiguity \cite{nozaki2022addressing} or tokenization inconsistency \cite{yan2025addressing}. Figure~\ref{fig:pipeline} contrasts sender-side oracle evaluation with receiver-realistic communication. The former reuses the generation-time token sequence, whereas the latter reconstructs tokens from the transmitted stegotext. A local mismatch can alter subsequent coding states and corrupt the remaining secret bits \cite{wang2026retoksync}. Consequently, evaluations that assume preserved generation-time tokenization can substantially overstate practical recovery.

Prior work has addressed tokenization inconsistency mainly through inference-time disambiguation. Segmentation Tricks (SegTrick) combines stepwise tokenization with prefix-based token disambiguation \cite{nozaki2022addressing}; Stepwise Verification removes candidates that would yield inconsistent retokenization \cite{yan2025addressing}; SyncPool groups prefix-related candidates to synchronize sampling \cite{qi2025provably}. These methods operate on candidate sets induced by a fixed model distribution. Although they can prevent or contain decoding desynchronization, they do not adapt the generator toward sequences that remain stable after retokenization. Moreover, tokenization consistency alone does not ensure fluent or statistically inconspicuous stegotext. Inference-time disambiguation therefore does not jointly optimize receiver-side recoverability, linguistic quality, and anti-steganalysis performance.

To address these limitations, we introduce TI-StegoAlign, which jointly optimizes generation for receiver-side recoverability after retokenization, linguistic quality, and anti-steganalysis performance. It treats the detokenization-retokenization cycle as part of the communication channel. Within this framework, we propose the Bit-Consistent Supervised Objective (BCSO), which couples secret information embedding with conditional generation while preserving in-domain language modeling capability. We further devise Channel-Conditioned Preference Optimization (CCPO), which learns channel-aware sequence preferences under a unified recovery–quality–security criterion. Together, these objectives adapt the generator itself to channel effects rather than correcting individual outputs at inference time.

Our main contributions are summarized below:

\begin{itemize}
    \item We establish a receiver-realistic covert communication setting in which secret recovery relies solely on the transmitted stegotext. Across 29{,}979 transmissions, 45.3\% of TI cases produce at least three additional bit errors after the first decoding error, exposing a reliability gap overlooked by sender-side oracle evaluation.

    \item We introduce TI-StegoAlign, a channel-guided policy alignment framework that adapts the generation policy to the detokenization-retokenization channel. Its Bit-Consistent Supervised Objective enlarges local coding margins at realized sender-side decisions while preserving in-domain generation capability.
    
    \item We further introduce Channel-Conditioned Preference Optimization, which learns channel-aware sequence preferences to jointly optimize receiver-side recoverability, text quality, and anti-steganalysis performance.
\end{itemize}

\begin{figure*}[t]
    \centering
    \includegraphics[width=1\linewidth]{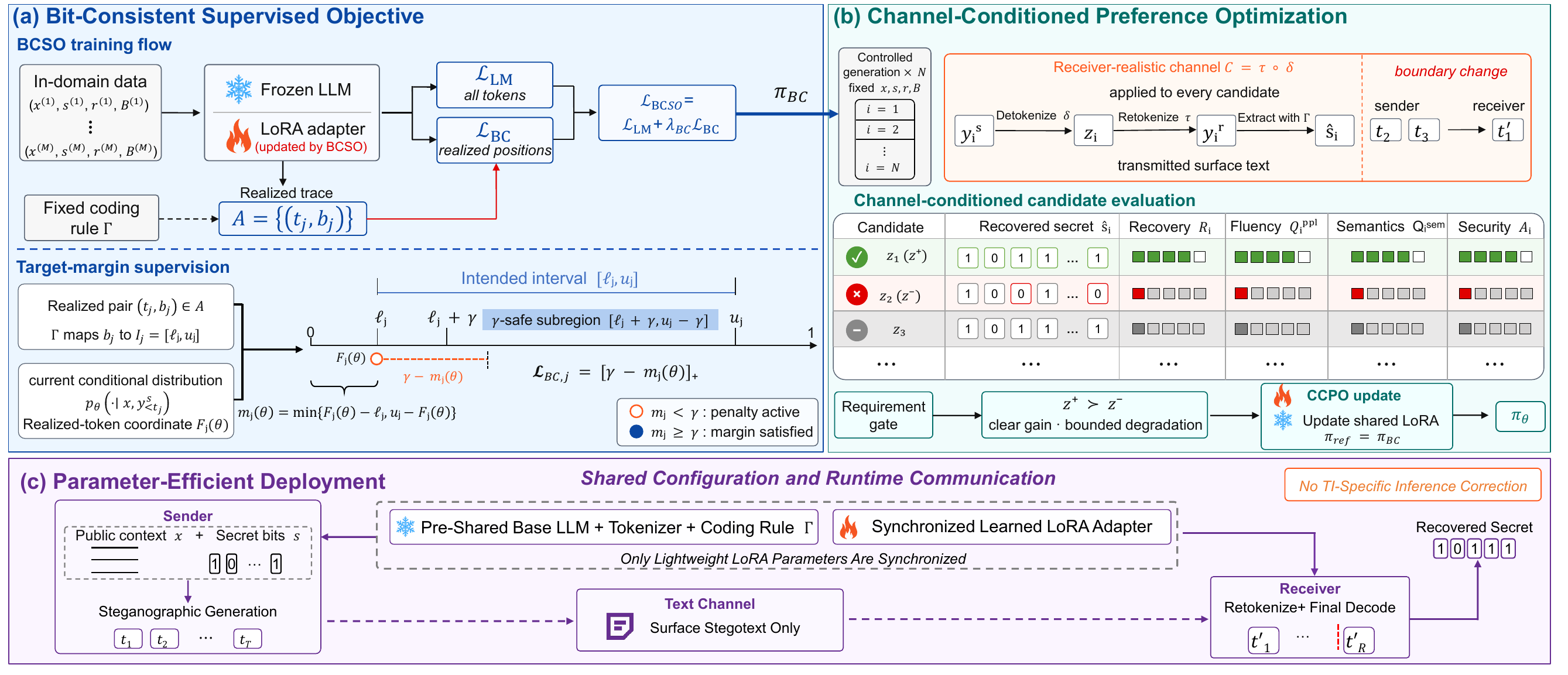}
    \caption{TI-StegoAlign framework: (a) BCSO combines in-domain language modeling with coding-margin supervision. (b) CCPO ranks complete stegotexts using receiver-realistic recovery, linguistic quality, semantic preservation, and steganalytic security. (c) Training updates only a shared LoRA adapter; communication transmits surface text without TI-specific correction.}
    \label{fig:framework}
\end{figure*}

\section{Related Work}

\subsection{Agent Covert Communication}
Existing work on covert communication among LLM agents has examined the risks and emergence of steganographic collusion and the design of explicit covert mechanisms. Motwani et al. \shortcite{motwani2024secret} formalize the threat model, capacity, and incentives of secret collusion among communicating agents, whereas Mathew et al. \shortcite{mathew2025hidden} demonstrate that steganographic protocols can emerge under misspecified objectives and resist standard monitoring and mitigation. CoMet \cite{xu2025comet} develops metaphor-mediated communication for multi-agent language games, while Whispering Agents \cite{huang2026whispering} constructs event-driven protocols from storage, timing, and behavioral signals. Together, these studies characterize covert agent communication at the levels of incentives, emergent behavior, and protocol design.

\subsection{Generative Text Steganography}
Generative text steganography embeds secret information through controlled choices during text generation. Early neural approaches combined language models with fixed-length, Huffman, or arithmetic coding \cite{yang2019rnnstega,ziegler2019neural,shen2020near}, followed by adaptive grouping for distribution-aware embedding \cite{zhang2021provably}. Subsequent studies developed distribution-preserving security mechanisms \cite{kaptchuk2021meteor,schroederdewitt2023perfectly,ding2023discop}, improved coding efficiency \cite{wang2025sparsamp,yan2026range}, and extended provable security to public-key settings \cite{zhang2025publickey}. Recent LLM-based methods have explored black-box embedding \cite{wu2024generative}, adaptive strategy evolution \cite{zhou2026auto}, and optimized generation distributions \cite{huang2026odstega}.

A reliability issue in generative text steganography arises when secret information is encoded through token selection: receiver-side retokenization may alter the sampled sequence and disrupt extraction. Segmentation Tricks remove prefix-ambiguous candidates \cite{nozaki2022addressing}, SyncPool synchronizes sampling over prefix-related candidate groups \cite{qi2025provably}, and Stepwise Verification filters candidates that would retokenize inconsistently \cite{yan2025addressing}. Secure disambiguation \cite{yan2023secure},
verification-based methods~\cite{yan2024verification}, tokenization-free methods~\cite{yan2024tokenfree}, and corrective state resets~\cite{wang2026retoksync} provide alternative remedies. Although these methods reduce decoding failures, they enforce reliability through external filtering or synchronization, which can constrain generation rather than enable the model to directly produce stegotext that remains recoverable while preserving linguistic quality and anti-steganalysis performance.

\subsection{Post-Training for Generative Text Steganography}

Language-model post-training offers a general mechanism for shaping sequence-level generation through demonstrations and preferences. Supervised fine-tuning (SFT) learns desired conditional behavior from demonstrations \cite{ouyang2022training}, whereas Direct Preference Optimization (DPO) directly optimizes a language-model policy from preference pairs without an explicit reward model \cite{rafailov2023direct}. Existing steganographic studies have mainly focused on inducing or evaluating covert behavior \cite{mathew2025hidden,karpov2025steganographic}. In contrast, TI-StegoAlign applies supervised and preference-based post-training to generative text steganography under receiver-realistic channel conditions, jointly optimizing secret recovery, linguistic quality, and steganalytic security.

\section{The Proposed Method}
As illustrated in Fig.~\ref{fig:framework}, TI-StegoAlign integrates the post-training procedure with a receiver-realistic communication pipeline. Given a public context $x$ and a secret bitstream $\mathbf{s}=(s_1,\ldots,s_L)\in\{0,1\}^{L}$, BCSO first establishes a coding-aware policy through local sender-side margin supervision. CCPO then optimizes complete stegotexts using receiver-realistic recovery, quality, and security outcomes. Parameter-efficient post-training confines adaptation to lightweight low-rank adapters while keeping the shared base model and coding rule fixed.

\subsection{Receiver-Realistic Communication Model}
TI-StegoAlign models steganographic communication using the detokenized stegotext that actually traverses the channel, rather than the sender-side token trajectory used internally during generation. This separation yields an operationally faithful communication model in which secret recovery depends exclusively on text available to the receiver, without assuming access to sender-side token IDs.

Let $\pi_{\theta}$ denote the language-model policy adapted by TI-StegoAlign, and let $\Gamma$ denote the fixed CDF-partition rule shared by the sender and receiver. Given the conditional distribution $p_{\theta,t}$, $\Gamma$ determines whether step $t$ is used for embedding. At a selected step, it maps the next coding symbol $b$, consumed from $\mathbf{s}$, to a CDF interval $\mathcal{I}_{\Gamma,t}(b)$. The sender samples a token assigned to this interval and advances the secret pointer accordingly. The receiver recomputes the conditional distribution from its retokenized prefix and applies the same partition and position-selection rule. Under condition $(x,\mathbf{s})$, this process induces a sender-side token sequence

\begin{equation}
\label{eq:sender-generation}
    \mathbf{y}^{s}
    \sim
    P_{\theta,\Gamma}
    \!\left(\,\cdot \mid x,\mathbf{s}\right),
    \qquad
    \mathbf{y}^{s}
    =
    \left(y^{s}_{1},\ldots,y^{s}_{T}\right).
\end{equation}

Here, $P_{\theta,\Gamma}$ denotes steganographic generation under $\pi_{\theta}$ and $\Gamma$. The formulation targets CDF/interval-based coding rules, whose bit-specific interval boundaries are available during
training. 

The sender-side token sequence is not exposed through the communication channel. Let $\delta$ denote detokenization and $\tau$ denote tokenization of the received text. The transmitted stegotext $z$ and the token sequence reconstructed by the receiver are

\begin{equation}
\label{eq:text-channel}
    z
    =
    \delta\!\left(\mathbf{y}^{s}\right),
    \qquad
    \mathbf{y}^{r}
    =
    \tau(z)
    =
    \tau\!\left(\delta\!\left(\mathbf{y}^{s}\right)\right)
    \equiv
    \mathcal{C}\!\left(\mathbf{y}^{s}\right),
\end{equation}

where $\mathcal{C}=\tau\circ\delta$ denotes the
detokenization--retokenization channel. Since the two operations are not generally inverse, the reconstructed sequence may satisfy
$\mathbf{y}^{r}\neq\mathbf{y}^{s}$ even when both parties use the same tokenizer. A local boundary change alters the prefix used to reconstruct the
next-token distribution. It may therefore change the position-selection decision or the interval membership of a later token. Once the secret pointers diverge, errors can propagate through subsequent extraction.

Secret recovery is therefore defined exclusively on the sequence reconstructed at the receiver:

\begin{equation}
\label{eq:receiver-recovery}
    \hat{\mathbf{s}}
    =
    \operatorname{Ext}_{\Gamma}
    \!\left(x,\mathbf{y}^{r};\pi_{\theta}\right)
    =
    \operatorname{Ext}_{\Gamma}
    \!\left(
        x,\mathcal{C}(\mathbf{y}^{s});\pi_{\theta}
    \right),
\end{equation}

where $\operatorname{Ext}_{\Gamma}$ denotes the extraction procedure defined by the shared coding rule $\Gamma$ and the conditional distributions produced by $\pi_{\theta}$. Recovery from $\mathbf{y}^{s}$ serves only as a sender-side oracle, whereas communication reliability is evaluated on the receiver-reconstructed sequence $\mathbf{y}^{r}$.

This channel model defines the communication conditions under which the learned generation policy is optimized. On this basis, TI-StegoAlign internalizes post-channel recoverability into the generation policy rather than treating it as an inference-time correction.

\subsection{Bit-Consistent Supervised Objective}
Bit-Consistent Supervised Objective (BCSO) initializes TI-StegoAlign with a coding-aware policy. Standard SFT maximizes the likelihood of in-domain demonstration tokens but does not account for their positions within the coding intervals induced by $\Gamma$. BCSO augments the language-modeling objective with local margin supervision at realized sender-side embedding positions, encouraging larger coding margins while preserving in-domain generation. Moving a realized token away from the nearest interval boundary reduces the sensitivity of its coding assignment to small shifts in the CDF coordinate. BCSO operates on fixed sender-side traces generated by the pre-BCSO policy under $\Gamma$. The token sequence, consumed coding symbols, and realized embedding positions remain fixed throughout supervised optimization.

To preserve in-domain generation, we retain the standard causal language-modeling loss over the sender-side sequence $\mathbf{y}^{s}$ introduced in Eq.~\eqref{eq:sender-generation}:
\begin{equation}
\label{eq:lm-objective}
\mathcal{L}_{\mathrm{LM}}
=
-\frac{1}{T}
\sum_{t=1}^{T}
\log p_{\theta}\!\left(
y^{s}_{t}
\mid
x,\mathbf{y}^{s}_{<t}
\right).
\end{equation}
This term preserves the model's conditional generation capability in the target domain. Let $\mathcal{A}=\{(t_j,b_j)\}_{j=1}^{|\mathcal{A}|}$ denote the realized embedding trace, where $b_j$ is the coding symbol consumed
from the secret bitstream at position $t_j$. Under the current coding state, $\Gamma$ maps $b_j$ to its intended CDF interval $\mathcal{I}_j=[\ell_j,u_j]$. Consistent with the shared coding rule, a token is assigned to the interval containing its upper CDF endpoint; this convention also resolves cases in which its probability mass crosses an interval boundary. Let $\sigma_{t_j}$ denote the descending probability ordering and $k_j$ the rank of the realized
token. The corresponding endpoint is

\begin{equation}
\label{eq:current-cumulative-coordinate}
F_j(\theta)
=
\sum_{i=1}^{k_j}
p_{\theta}\!\left(
v_{\sigma_{t_j}(i)}
\mid x,\mathbf{y}^{s}_{<t_j}
\right).
\end{equation}

We define its coding margin as the distance to the nearest boundary of the intended interval:

\begin{equation}
\label{eq:coding-margin}
m_j(\theta)
=
\min\left\{
F_j(\theta)-\ell_j,\,
u_j-F_j(\theta)
\right\}.
\end{equation}

The margin is positive when the realized token falls within its intended interval and increases with its distance from the nearest interval boundary. At each forward pass, the current policy recomputes the probability ordering, realized token rank, and interval assignment. These discrete quantities are held fixed during backpropagation, so gradients propagate only through the cumulative probabilities defining $F_j(\theta)$.

The bit-consistency loss and the complete supervised objective are
\begin{equation}
\label{eq:bcso-objective}
\begin{aligned}
\mathcal{L}_{\mathrm{BC}}
&=
\frac{1}{|\mathcal{A}|}
\sum_{(t_j,b_j)\in\mathcal{A}}
\left[
\gamma-m_j(\theta)
\right]_{+},\\
\mathcal{L}_{\mathrm{BCSO}}
&=
\mathcal{L}_{\mathrm{LM}}
+
\lambda_{\mathrm{BC}}
\mathcal{L}_{\mathrm{BC}},
\end{aligned}
\end{equation}
where $[a]_{+}=\max(a,0)$, $\gamma>0$ is chosen such that $2\gamma<u_j-\ell_j$ for every realized interval, and $\lambda_{\mathrm{BC}}$ controls the strength of bit-consistency supervision.
$\mathcal{L}_{\mathrm{LM}}$ preserves in-domain generation, whereas $\mathcal{L}_{\mathrm{BC}}$ moves realized-token coordinates toward the interior of their intended intervals. The resulting policy initializes the receiver-conditioned optimization in CCPO.

\subsection{Channel-Conditioned Preference Optimization}
\label{CCPO}
Building on the BCSO policy, Channel-Conditioned Preference Optimization (CCPO) further aligns generation with the outcomes of receiver-realistic communication. CCPO constructs preferences from the recoverability, linguistic quality, and steganalytic security observed after the generated stegotext traverses the detokenization-retokenization channel.

Let $\pi_{\mathrm{BC}}$ denote the policy obtained through BCSO. For each training instance, CCPO samples $N$ sender-side token sequences $\mathbf{y}^{s}_{i}$ under the same public context $x$, secret bitstream $\mathbf{s}$, reference text $r$, and generation budget $B$. Here, $r$ is the non-steganographic reference text used to assess semantic
preservation, and $B$ is the token-generation budget. Only the sampling randomness varies.

Each candidate is characterized by
$\mathbf{u}_i=(R_i,Q_i^{\mathrm{ppl}},Q_i^{\mathrm{sem}},A_i)$. Here, $R_i$ measures receiver-side recovery, $Q_i^{\mathrm{ppl}}$ measures fluency relative to the target-domain distribution, and $Q_i^{\mathrm{sem}}$ measures semantic consistency. The steganalytic security score $A_i$ is computed by a frozen evaluator that combines a domain-specific BERT classifier with fixed stylistic statistics. The classifier is trained separately for each domain using balanced cover-stego pairs from the corresponding training split. No validation or test text is used for training. Exact score definitions and evaluator details are provided in Appendix~A.

Preference construction first removes candidates that violate minimum requirements for receiver-side recovery, fluency, semantic consistency, steganalytic security, or length. The remaining candidates are ordered by a fixed recovery-dominant score $S_i$ computed from the four components of $\mathbf{u}_i$, with deterministic tie-breaking. Only an ordered pair with $S_i>S_j$ is tested under the pairwise preference criterion:

\begin{equation}
\label{eq:margin-constrained-preference}
z_i \succ z_j
\Longleftrightarrow
\begin{cases}
S_i>S_j,\\
\exists k:\; u_{i,k} \geq u_{j,k}+\Delta_k,\\
\forall \ell\neq k:\; u_{i,\ell}
\geq u_{j,\ell}-\varepsilon_{\ell\mid k},
\end{cases}
\end{equation}
where $\Delta_k>0$ denotes the required improvement on criterion $k$, and $\varepsilon_{\ell\mid k}\geq0$ bounds the admissible degradation on criterion $\ell$ when the preference is established along axis $k$. The score orients candidate pairs but is insufficient by itself; a pair is retained only when the conditions also hold. Since every retained pair follows the same strict score order, reciprocal preferences and within-set cycles are excluded. Exact score weights, feasibility thresholds, margins, tolerances, diversity filters, and tie-breaking rules are reported in Appendix~A. All candidates in a preference set are generated under the same $(x,\mathbf{s},r,B)$, and their labels are determined from receiver-realistic outcomes. DPO then scores the canonical tokenizations $\widetilde{\mathbf{y}}^{\pm}=\tau(z^{\pm})$ under a shared textual prompt $c$ containing $x$, $r$, and the bit-length instruction. The secret and coding state are not included in $c$; instead, they affect pair construction through $\Gamma$-constrained generation and receiver-side extraction. Hence,
$\mathcal{D}_{\mathrm{CC}}
=
\{(c,\widetilde{\mathbf{y}}^{+},\widetilde{\mathbf{y}}^{-})
\mid z^{+}\succ z^{-}\}$. With $\pi_{\mathrm{ref}}=\pi_{\mathrm{BC}}$, define
\begin{equation}
\label{eq:delta-c-y}
\Delta_{\theta}^{\mathrm{LM}}(c,\mathbf{y})
=
\log \pi_{\theta}(\mathbf{y}\mid c)
-
\log \pi_{\mathrm{ref}}(\mathbf{y}\mid c).
\end{equation}
Following DPO \cite{rafailov2023direct}, the preference pairs are optimized
using
\begin{equation}
\label{eq:ccpo-objective}
\mathcal{L}_{\mathrm{CCPO}}
=
-\mathbb{E}_{\mathcal{D}_{\mathrm{CC}}}
\left[
\log \sigma\!\left(
\beta
\left[
\Delta_{\theta}^{\mathrm{LM}}(c,\widetilde{\mathbf{y}}^{+})
-
\Delta_{\theta}^{\mathrm{LM}}(c,\widetilde{\mathbf{y}}^{-})
\right]
\right)
\right],
\end{equation}
where $\beta$ controls deviation from the supervised policy. Accordingly, CCPO is a channel-observed preference optimization rather than exact DPO over $P_{\theta,\Gamma}$: receiver-side outcomes determine the labels, while gradients update the LLM probabilities subsequently partitioned by $\Gamma$.

CCPO converts receiver-realistic outcomes for secret recovery, linguistic quality, semantic preservation, and steganalytic security into pairwise preferences over complete stegotexts. Together, BCSO and CCPO form a unified post-training framework: BCSO establishes a coding-aware policy through supervised optimization, while CCPO further aligns it with these stegotext preferences.

\subsection{Parameter-Efficient Training and Communication}

BCSO and CCPO sequentially update a shared set of LoRA adapters \cite{hu2022lora}, while the base language model, tokenizer, and coding rule $\Gamma$ remain fixed. The resulting policy is shared by the sender and receiver during communication. The sender transmits only the detokenized stegotext $z$, which the receiver retokenizes and extracts under $\Gamma$. This deployment requires no TI-specific candidate filtering, path verification, or decoding repair at inference time.

\section{Experiments}

This section presents the experimental setup and results. Section \ref{setup} details the experimental setup. Section~\ref{communication} assesses secret recovery under receiver-realistic communication. Sections~\ref{imperceptibility} and~\ref{steganalysis} evaluate imperceptibility and anti-steganalysis performance, respectively. Section~\ref{human-evaluation} reports the human evaluation results. Appendix~C presents fine-grained results for each domain. Appendix~D reports the ablation studies, Appendix~E analyzes runtime efficiency, and Appendix~F provides qualitative examples.

\subsection{Experimental Settings}\label{setup}

\begin{table*}[t]
\centering
\small
\renewcommand{\arraystretch}{1.08}
\setlength{\tabcolsep}{4pt}
\begin{tabular*}{\textwidth}{
@{\extracolsep{\fill}}
lcccccc
@{}
}
\toprule
Method &
\shortstack{Oracle Bit\\Acc. (\%) $\uparrow$} &
\shortstack{Receiver Bit\\Acc. (\%) $\uparrow$} &
\shortstack{Oracle-to-Receiver\\Gap $\downarrow$} &
\shortstack{Receiver Exact Message\\Recovery (\%) $\uparrow$} &
\shortstack{TI Rate\\(\%) $\downarrow$} &
\shortstack{Cascading Error\\Rate (\%) $\downarrow$} \\
\midrule
ADG
& 99.94 & 77.31 & 22.63 & 72.91 & 24.47 & 14.49 \\

Discop
& 98.60 & 83.17 & 15.43 & 72.83 & 16.17 & 16.42 \\

SegTrick
& \textbf{100.00}
& \textbf{100.00}
& \textbf{0.00}
& \textbf{100.00}
& 0.86
& \textbf{0.00} \\

SyncPool
& \textbf{100.00} & \textbf{100.00} & \textbf{0.00} & \textbf{100.00} & 0.97 & \textbf{0.00} \\

StepVerify
& \textbf{100.00} & \textbf{100.00} & \textbf{0.00} & \textbf{100.00} & 0.78 & \textbf{0.00} \\

\textbf{TI-StegoAlign}
& \textbf{100.00}
& \textbf{100.00}
& \textbf{0.00}
& \textbf{100.00}
& \textbf{0.76}
& \textbf{0.00} \\
\bottomrule
\end{tabular*}
\caption{Comparison of communication reliability across methods. ($\uparrow$ higher is better, $\downarrow$ lower is better).}
\label{tab:communication-reliability}
\end{table*}

\textbf{Datasets.} We evaluate TI-StegoAlign on three publicly available corpora: the News Category Dataset (News) \cite{misra2022news}, the Large Movie Review Dataset (Movie) \cite{maas2011learning}, and Sentiment140 (Tweet) \cite{go2009twitter}. The evaluation set contains 12{,}300 News samples, with 300 samples drawn from each of 41 selected categories, together with 20{,}000 Movie reviews and 20{,}000 tweets. All methods considered in the evaluation are applied to the full set of 52{,}300 instances, resulting in 313{,}800 transmissions across the six methods. The three corpora span diverse domains, writing styles, and sequence lengths. Results in the main text are averaged across News, Movie, and Tweet, while domain-specific results are reported in Appendix~C. The auxiliary channel characterization in Section~\ref{communication} is conducted on an independently collected set of 29{,}979 transmissions, which is excluded from the count above.

\textbf{Baselines.} We compare TI-StegoAlign with five representative baselines to support a comprehensive evaluation across the considered dimensions. ADG \cite{zhang2021provably} and Discop \cite{ding2023discop} are representative distribution-aware generative steganography methods. Segmentation Tricks \cite{nozaki2022addressing}, SyncPool \cite{qi2025provably}, and Stepwise Verification \cite{yan2025addressing} explicitly address segmentation ambiguity or tokenization inconsistency through generation-time disambiguation. For brevity, we refer to Segmentation Tricks and Stepwise Verification as SegTrick and StepVerify, respectively. For a fair comparison, all methods use the same underlying language model and tokenizer whenever applicable and are evaluated at a common target payload of 0.5 bpw. The results therefore represent controlled re-evaluations rather than reproductions of the original operating points; method-specific settings and results are reported in Appendix~G.

\textbf{Implementation Details.} We use Llama-3.1-8B \cite{grattafiori2024llama} as the base language model together with its corresponding tokenizer. To further assess TI-StegoAlign across different model backbones, we additionally conduct experiments with Qwen3-8B \cite{yang2025qwen3}. The corresponding results are reported in Appendix~H. LoRA adapters are inserted into the query, key, value, and output projections of each self-attention module, with rank $r=16$, scaling factor $\alpha=32$, and dropout 0.05; all other model parameters remain frozen. The policy obtained after BCSO is used to initialize CCPO. Complete optimization hyperparameters, generation settings, and hardware specifications are provided in Appendix~B. Statistical tests are selected by metric. We use two-sided exact McNemar tests for paired exact-recovery outcomes, paired Wilcoxon signed-rank tests \cite{wilcoxon1945individual} for per-instance $\mathrm{PPL}^{*}$ and SS, and paired bootstrap tests with 95\% confidence intervals for differences in bit and detector accuracy. KLD is reported descriptively. Holm's procedure \cite{holm1979simple} is applied within each metric family at $\alpha=0.05$. All reported pairwise improvements remain significant after correction.

\begin{table}[t]
\centering
\small
\renewcommand{\arraystretch}{1.08}
\setlength{\tabcolsep}{2.5pt}
\begin{tabular*}{\columnwidth}{
@{\extracolsep{\fill}}
lccc
@{}
}
\toprule
Method &
$\mathrm{PPL}^{*}\downarrow$ &
SS$\uparrow$ &
 KLD$\downarrow$ \\
\midrule
ADG
& 0.547
& 0.504
& 2.371 \\

Discop
& 0.756
& 0.502
& 2.331 \\

SegTrick
& 0.921
& 0.554
& 2.511 \\

SyncPool
& 0.774
& 0.508
& 2.775 \\

StepVerify
& 0.719
& 0.500
& 2.436 \\

\textbf{TI-StegoAlign}
& \textbf{0.429}
& \textbf{0.563}
& \textbf{1.687} \\
\bottomrule
\end{tabular*}
\caption{Comparison of imperceptibility across methods.}
\label{tab:imperceptibility}
\end{table}

\subsection{Communication Reliability}
\label{communication}

A covert channel is operationally reliable only when the intended secret can be recovered from the transmitted text available to the receiver. In our evaluation, Oracle Bit Accuracy measures recovery on the token sequence produced during generation, whereas Receiver Bit Accuracy is computed after the transmitted stegotext is retokenized at the receiver. The Oracle-to-Receiver Gap is the absolute difference between the two. Exact Message Recovery counts a sample as correct only when the complete secret is recovered. TI Rate reports the proportion of samples whose token sequence changes after detokenization and retokenization. Cascading Error Rate measures the bit-error rate over the remaining secret positions following the first decoding error and is set to zero when no decoding error occurs. For the channel analysis, we additionally report Cascade Incidence, defined as the proportion of TI cases in which at least three additional secret bits are decoded incorrectly after the first decoding error. Results at method-specific embedding rates are provided in Appendix~G.

We establish a receiver-realistic covert communication setting in which secret recovery relies solely on the transmitted stegotext. In a separate channel-characterization study of 29{,}979 transmissions, 45.3\% of transmissions exhibiting TI produce at least three additional bit errors after the first decoding error. This result shows that a local tokenization mismatch can propagate through subsequent coding states. As shown in Table~\ref{tab:communication-reliability}, ADG and Discop retain high Oracle Bit Accuracy values of 99.94\% and 98.60\%, respectively, but their Receiver Bit Accuracy decreases to 77.31\% and 83.17\% after transmission, producing Oracle-to-Receiver Gaps of 22.63 and 15.43 percentage points. These results confirm that sender-side recovery can substantially overestimate reliability over the surface-text channel. By contrast, SegTrick, SyncPool, StepVerify, and TI-StegoAlign all achieve 100.00\% Oracle and Receiver Bit Accuracy, 100.00\% Exact Message Recovery, and zero Oracle-to-Receiver Gap and cascading errors. Among these methods, TI-StegoAlign records the lowest TI Rate of 0.76\%. A nonzero TI rate remains compatible with exact recovery because a changed token boundary does not necessarily alter subsequent interval assignments or the secret-pointer state. TI-StegoAlign achieves perfect recovery without modifying candidate pools or introducing additional synchronization or verification procedures at inference time. Practical covert communication requires more than reliable secret recovery. The transmitted stegotext must also conceal the presence of embedded information by remaining linguistically natural and statistically inconspicuous. Therefore, the following section evaluates imperceptibility.

\subsection{Imperceptibility}
\label{imperceptibility}

Imperceptibility in text steganography is commonly assessed from two perspectives: text quality and statistical imperceptibility. Text quality evaluates fluency and consistency with a paired reference text,
while statistical imperceptibility measures the distributional similarity
between the non-steganographic cover corpus and the stego corpus.

\textbf{Text Quality.}
High quality stegotext should remain linguistically natural while preserving the semantics of its corresponding cover text. Fluency is evaluated using perplexity (PPL) \cite{mikolov2010recurrent}, computed with GPT-2 model. To compare across corpora with different baseline perplexities, we report the normalized perplexity deviation $\mathrm{PPL}^{\ast}=\frac{\lvert \mathrm{PPL}_{\text{stego}}-\mathrm{PPL}_{\text{cover}}\rvert}{\mathrm{PPL}_{\text{cover}}}$,
where a lower value indicates closer agreement with the domain-specific cover distribution~\cite{zhou2026auto}. Semantic similarity (SS) is computed as the cosine similarity between sentence embeddings of a stegotext and its paired reference text $r$. For each test instance, $r$ is the original corpus text and is shared by all methods. We compute embeddings with the RoBERTa-based roberta-base-nli-mean-tokens Sentence-BERT model \cite{reimers2019sentence,liu2019roberta}.

As shown in Table~\ref{tab:imperceptibility}, TI-StegoAlign achieves the best results on both text-quality metrics, with a $\mathrm{PPL}^{*}$ of 0.429 and an SS of 0.563. Compared with the strongest baseline on each metric, it reduces $\mathrm{PPL}^{*}$ by 21.6\% relative to ADG and improves SS by 0.009 over SegTrick. Together with the reliability results in Section~\ref{communication}, these findings show that TI-StegoAlign achieves the same perfect-recovery as the disambiguation baselines while providing better fluency and semantic fidelity.

\textbf{Statistical Imperceptibility.}
We use Kullback-Leibler divergence (KLD) \cite{cachin1998information,zhang2021provably} to quantify the distributional discrepancy between cover and stego texts, where a lower value indicates stronger statistical imperceptibility. In our experiments, we take the logarithm of the KLD values to facilitate a clearer comparison of statistical imperceptibility between algorithms. As shown in Table~\ref{tab:imperceptibility}, TI-StegoAlign obtains a KLD of 1.687, reducing the strongest baseline, Discop, by 0.644. This result indicates that the learned generation policy remains closely aligned with the cover-text distribution while maintaining exact receiver-side recovery under receiver-realistic communication.

The improvements in linguistic and statistical imperceptibility are consistent with CCPO's multi-criteria preference construction. By jointly considering receiver-side recovery, linguistic quality, semantic preservation, and steganalytic security, CCPO guides policy optimization through preferences that reflect trade-offs among these criteria rather than any single score.

\begin{table}[t]
\centering
\small
\renewcommand{\arraystretch}{1.08}
\setlength{\tabcolsep}{2.5pt}
\begin{tabular*}{\columnwidth}{
@{\extracolsep{\fill}}
lcccc
@{}
}
\toprule
Method &
BD (\%)&
BF (\%)&
SANet (\%)&
Average (\%)\\
\midrule
ADG
& 71.73
& 64.97
& \textbf{57.33}
& 64.68 \\

Discop
& 69.17
& 79.17
& 73.06
& 73.80 \\

SegTrick
& 68.00
& 79.67
& 68.79
& 72.15 \\

SyncPool
& 77.67
& 69.17
& 74.44
& 73.76 \\

StepVerify
& 64.16
& 69.83
& 71.49
& 68.49 \\

\textbf{TI-StegoAlign}
& \textbf{51.00}
& \textbf{62.07}
& 68.75
& \textbf{60.61} \\
\bottomrule
\end{tabular*}
\caption{Comparison of anti-steganalysis performance across methods. Average denotes the mean over BD, BF, and SANet.}
\label{tab:steganalysis}
\end{table}

\begin{figure}[t]
  \centering
  \includegraphics[width=0.95\linewidth]{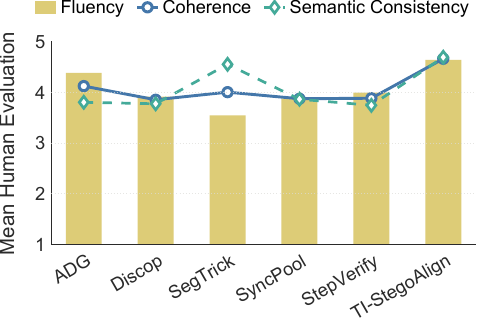}
  \caption{The results of the human evaluation.}
  \label{human}
\end{figure}

\subsection{Security}
\label{steganalysis}

Steganography and steganalysis evolve through a continuing adversarial process: steganographic methods seek to conceal the traces of secret embedding, while steganalyzers attempt to identify them. Anti-steganalysis performance therefore provides an important measure of steganographic security. We evaluate anti-steganalysis performance using BiLSTM-Dense (BD) \cite{yang2020linguistic}, BERT-FT (BF) \cite{peng2021realtime}, and SANet \cite{xue2024sanet}. Notably, the BERT-based evaluator used in CCPO and the BF detector are trained independently and share no parameters. For each method and domain, we construct 10{,}000 cover-stego pairs for training and use 1{,}000 pairs each for validation and testing. Accuracy closer to 50\%, corresponding to random
guessing, indicates stronger anti-steganalysis performance.

As shown in Table~\ref{tab:steganalysis}, TI-StegoAlign achieves accuracies of 51.00\%, 62.07\%, and 68.75\% under BD, BF, and SANet, respectively. Its mean accuracy of 60.61\% is the closest to random guessing. TI-StegoAlign is closest to 50\% under BD and BF, whereas ADG remains stronger under SANet, with an accuracy of 57.33\%. The security gain is therefore broad but not uniform across detector architectures.

The absolute accuracies of ADG and Discop are not directly comparable with their original reports because the language models, payload regimes, steganalyzers, and evaluation protocols differ. Accordingly, the results here characterize their behavior under the shared Llama-3.1-8B benchmark rather than reproduce the original security claims. Overall, the lower average detectability is consistent with CCPO's use of steganalysis risk in channel-conditioned preference construction.

\subsection{Human Evaluation}
\label{human-evaluation}
To examine whether the improvements measured by metrics are reflected in human judgments, we conduct a blinded evaluation of fluency, coherence, and semantic consistency. For each method, we randomly sample 50 outputs from each domain using the same test instances and embedding rate. Each item contains the reference text and an anonymized stegotext, with method identities concealed and presentation order randomized. Ten evaluators who were not involved in model development independently rate each output on a five-point Likert scale. Fluency assesses grammaticality and naturalness, coherence measures logical continuity, and semantic consistency evaluates preservation of the reference content. Each output is rated by all evaluators. We report Krippendorff's alpha \cite{krippendorff2011alpha} and assess method effects with a cumulative-link mixed model containing evaluator and test-instance random effects. The complete rubric and statistical results are provided in Appendix~I.

As shown in Fig.~\ref{human}, TI-StegoAlign receives the highest human ratings across fluency, coherence, and semantic consistency. The judgments agree with the text-quality results, indicating that its stegotext remains natural, coherent, and faithful to the paired reference. These qualities preserve the communicative function of the transmitted text while conveying hidden information.

\section{Conclusion}
This work formulates tokenization inconsistency as a receiver-side channel problem and introduces TI-StegoAlign, a channel-aware post-training framework for reliable generative text steganography. BCSO enables bit-consistent secret embedding while preserving the base model's in-domain generation capability. CCPO further aligns the policy with receiver-realistic preferences over recovery,
linguistic quality, and anti-steganalysis performance. Both stages operate on the same frozen base model and update only a shared LoRA adapter. Experiments across diverse domains show that TI-StegoAlign achieves 100\% receiver-side bit accuracy and exact message recovery while improving text quality and anti-steganalysis performance.

\bibliography{aaai2027}

\clearpage
\section*{Appendix}

This appendix provides the methodological and experimental details that support the main paper. Appendix~\ref{app:ccpo} specifies CCPO candidate evaluation and preference construction. Appendix~\ref{app:implementation}
describes the data pipeline, evaluation metrics, statistical analysis, and
implementation settings. The remaining appendices report domain-specific
results, ablations, efficiency measurements, qualitative analyses,
embedding-rate studies, cross-backbone results, and the full human-evaluation
protocol. Unless stated otherwise, experiments follow the setup described in
Section~4.1 of the main paper.

\appendix

\section{CCPO Candidate Evaluation and Preference Construction}
\label{app:ccpo}

Section~3.3 of the main paper introduces CCPO using receiver-realistic
communication outcomes. This appendix specifies the candidate measures,
screening criteria, and pair-construction procedure used to instantiate that
objective.

\subsection{Controlled Candidate Generation}
For each training condition $(x,\mathbf{s},r,B)$, the BCSO policy
$\pi_{\mathrm{BC}}$ samples $N_d$ candidate stegotexts, where $d$ denotes the
text domain. The public context $x$, secret bitstream $\mathbf{s}$, semantic
reference $r$, and generation budget $B$ remain fixed within the candidate
set; only the sampling randomness varies. We use $N_{\mathrm{News}}=8$ and
$N_{\mathrm{Movie}}=N_{\mathrm{Tweet}}=16$. The sampling parameters and
length controls are reported in Appendix~\ref{app:implementation}.

Each sender-side sequence is evaluated after passing through the same
receiver-realistic channel used in the main experiments:
\begin{equation}
\mathbf{y}^{\mathrm{s}}_i
\xrightarrow{\delta} z_i
\xrightarrow{\tau} \mathbf{y}^{\mathrm{r}}_i
\xrightarrow{\operatorname{Ext}_{\Gamma}}
\hat{\mathbf{s}}_i.
\end{equation}
Thus, preference labels are determined from the transmitted stegotext $z_i$
and the sequence reconstructed by the receiver.

\subsection{Candidate Evaluation}
Each candidate is represented by
\begin{equation}
\mathbf{u}_i
=
\bigl(R_i,Q_i^{\mathrm{ppl}},Q_i^{\mathrm{sem}},A_i\bigr),
\end{equation}
where larger values are preferred. Receiver-side recovery is
\begin{equation}
R_i
=
1-\frac{e(\mathbf{s},\hat{\mathbf{s}}_i)}{|\mathbf{s}|},
\end{equation}
with missing decoded positions counted as errors. Let $\mu_{\mathrm{ppl},d}$
denote the reference perplexity of domain $d$. We define
\begin{equation}
d_i^{\mathrm{ppl}}
=
\left|\mathrm{PPL}(z_i)-\mu_{\mathrm{ppl},d}\right|,
\qquad
Q_i^{\mathrm{ppl}}=-d_i^{\mathrm{ppl}},
\end{equation}
using reference values 108.58, 113.87, and 867.01 for News, Movie, and Tweet,
respectively. Semantic consistency is measured by
\begin{equation}
Q_i^{\mathrm{sem}}
=
\cos\!\bigl(h(z_i),h(r)\bigr),
\end{equation}
where $h$ is the sentence encoder described in Appendix~\ref{app:implementation}.

The steganalytic security score is $A_i=1-D_i$, where $D_i$ is the estimated
probability that $z_i$ is steganographic. The estimate combines the output of
a domain-specific BERT-based classifier with fixed stylistic statistics. The
classifier is initialized from BERT-base and trained
separately for each domain on balanced sets of cover and stego texts drawn
exclusively from the corresponding training split. The stylistic component
uses sentence-length variation, type-token ratio, and punctuation density.
All evaluator components are fixed before CCPO, and no validation or test text
is used to train the classifier. The BERT-based evaluator used here is trained
independently of the BF detector used in the final security evaluation.

\subsection{Candidate Screening and Diversity Control}
Candidates must satisfy $R_i\geq0.95$ and $A_i\geq0.30$. After feasibility screening and diversity filtering, eligible candidates are
ranked by $S_i$. We retain the two highest- and two lowest-scoring candidates
for News, and the three highest- and three lowest-scoring candidates for
Movie and Tweet. These subsets determine which candidate pairs are examined,
but do not by themselves establish the final preference. The remaining requirements are domain-specific and are summarized in Table~\ref{tab:ccpo-screening}. Word-count deviation is measured relative to the corresponding training-domain mean.


\begin{table}[t]
\centering
\small
\setlength{\tabcolsep}{12pt}
\begin{tabular}{@{}lcccc@{}}
\toprule
Domain & $N_d$ & PPL interval & $Q_{\min}^{\mathrm{sem}}$
& $\Delta_w^{\max}$ \\
\midrule
News  & 8  & $[60,150]$  & 0.65 & 8  \\
Movie & 16 & $[40,200]$  & 0.55 & 10 \\
Tweet & 16 & $[100,3000]$ & 0.52 & 5  \\
\bottomrule
\end{tabular}
\caption{Domain-specific candidate-generation and screening settings.
$\Delta_w^{\max}$ denotes the maximum permitted word-count deviation.}
\label{tab:ccpo-screening}
\end{table}

Exact duplicate texts are removed before ranking. A deterministic character-level similarity filter based on shared matching
subsequences then prevents near-duplicate candidates from entering the same pool;
the within-pool similarity threshold is 0.55, and candidate pairs with
similarity above 0.70 are discarded. These filters reduce redundant
preferences without affecting the evaluation criteria in $\mathbf{u}_i$.

\subsection{Preference Pair Construction}
Eligible candidates are ordered using
\begin{equation}
S_i
=
100R_i
-10\frac{d_i^{\mathrm{ppl}}}{\mu_{\mathrm{ppl},d}}
+5Q_i^{\mathrm{sem}}
+3A_i.
\label{eq:ccpo-orientation}
\end{equation}
The score provides a deterministic orientation for candidate pairs but does
not alone establish a preference. Ties are resolved by higher receiver-side
recovery, smaller perplexity deviation, higher semantic consistency, higher
steganalytic security, and finally the normalized text string, in that order.
The highest- and lowest-ranked eligible candidates form the preferred and
comparison pools shown in Table~\ref{tab:ccpo-screening}.

For an oriented pair with $S_i>S_j$, CCPO retains $z_i\succ z_j$ only when one
criterion improves by a prescribed amount and degradation in each remaining
criterion stays within its corresponding tolerance. The numerical settings are
reported compactly in Table~\ref{tab:ccpo-pairing}; domain-specific values are
ordered as News, Movie, and Tweet.

\begin{table*}[t]
\centering
\small
\begin{tabular}{lll}
\toprule
Preference axis & Required improvement & Maximum degradation in the remaining criteria \\
\midrule
Recovery
& $R_i-R_j\geq0.05$
& $d^{\mathrm{ppl}}$: 15; $Q^{\mathrm{sem}}$: 0.08; $A$: 0.10 \\
Fluency
& $d_j^{\mathrm{ppl}}-d_i^{\mathrm{ppl}}\geq 8/12/20$
& $R$: 0.05; $Q^{\mathrm{sem}}$: 0.02/0.03/0.03; $A$: 0.10 \\
Semantics
& $Q_i^{\mathrm{sem}}-Q_j^{\mathrm{sem}}\geq0.05/0.03/0.03$
& $R$: 0.05; $d^{\mathrm{ppl}}$: 5/8/15; $A$: 0.10 \\
Security
& $A_i-A_j\geq0.10$
& $R$: 0.05; $d^{\mathrm{ppl}}$: 15; $Q^{\mathrm{sem}}$: 0.08 \\
\bottomrule
\end{tabular}
\caption{CCPO pair-retention criteria. Domain-specific values are listed as News/Movie/Tweet.}
\label{tab:ccpo-pairing}
\end{table*}

All cross-pool combinations are tested using these conditions. A training
condition is omitted when no valid pair remains after screening, diversity
control, and pairwise comparison.

\subsection{Preference Data}
For each retained pair, CCPO stores the public prompt $c$ together with the
chosen and rejected surface texts. The prompt contains the public context,
semantic reference, and bit-length instruction, but not the secret-bit values
or coding state. Both texts are tokenized with the shared tokenizer, yielding
\begin{equation}
\mathcal{D}_{\mathrm{CC}}
=
\left\{
\bigl(c,\tau(z^+),\tau(z^-)\bigr)
\;\middle|\;
z^+\succ z^-
\right\}.
\end{equation}
The BCSO policy initializes both the trainable policy and the fixed reference
policy, so $\pi_{\mathrm{ref}}=\pi_{\mathrm{BC}}$ at the start of CCPO.

\section{Experimental and Reproducibility Details}
\label{app:implementation}

This appendix complements the experimental setup in Section~4.1 of the main
paper with the data-processing, metric, optimization, and computing details
needed to reproduce the reported results. Method-specific baseline settings
and embedding-rate configurations are provided in
Appendix~\ref{app:capacity}.

\subsection{Datasets and Data Splits}
The main evaluation contains 12,300 News samples, 20,000 Movie reviews, and
20,000 tweets. Applying all six methods to the same 52,300 instances produces
313,800 method--instance transmissions. A separate channel-characterization
study uses 29,979 independently collected transmissions and is not included in
this total.

All samples are drawn from the public corpora described in the main paper. For
News, 300 instances are randomly selected from each of 41 categories. Movie
and Tweet instances are randomly sampled from their corresponding corpora. A
fixed seed of 42 is used for randomized selection and data partitioning. Empty
records and exact duplicates are removed before the post-training data are
split into mutually exclusive training, validation, and test subsets using an
80/10/10 ratio. Table~\ref{tab:posttraining-splits} reports the resulting
counts. The original corpus text serves as the semantic reference $r$.

\begin{table}[t]
\centering
\small
\setlength{\tabcolsep}{8pt}
\begin{tabular}{lrrrr}
\toprule
Domain & Total & Train & Val. & Test \\
\midrule
News  & 12,300 & 9,840 & 1,230 & 1,230 \\
Movie & 20,000 & 16,000 & 2,000 & 2,000 \\
Tweet & 20,000 & 16,000 & 2,000 & 2,000 \\
\bottomrule
\end{tabular}
\caption{Post-training data splits.}
\label{tab:posttraining-splits}
\end{table}

\subsection{Generation Settings}
Generation follows the receiver-realistic protocol defined in the main paper.
Across all domains, sampling uses temperature 0.9, top-$k=50$, top-$p=0.92$,
and repetition penalty 1.05. News and Movie outputs are generated with 25--35
new tokens, whereas Tweet outputs use 10--25 new tokens to accommodate their
shorter form. The receiver always
retokenizes the transmitted surface text before extraction and never accesses
the sender-side token sequence.

\subsection{Communication-Reliability Metrics}
For instance $n$, let $\mathbf{s}_n\in\{0,1\}^{L_n}$ denote the intended
secret, $\hat{\mathbf{s}}_n^{\mathrm{o}}$ the sender-side oracle recovery, and
$\hat{\mathbf{s}}_n^{\mathrm{r}}$ the receiver recovery. We count each missing
or incorrectly decoded position as an error:
\begin{equation}
 e(\mathbf{s}_n,\hat{\mathbf{s}}_n)
 =
 \sum_{j=1}^{L_n}
 \mathbb{I}\!\left[
 j>|\hat{\mathbf{s}}_n|
 \;\lor\;
 \hat{s}_{n,j}\neq s_{n,j}
 \right].
\end{equation}
Oracle and Receiver Bit Accuracy are then
\begin{equation}
\mathrm{BA}^{q}
=
1-
\frac{\sum_{n=1}^{M}e(\mathbf{s}_n,\hat{\mathbf{s}}_n^{q})}
{\sum_{n=1}^{M}L_n},
\qquad q\in\{\mathrm{o},\mathrm{r}\}.
\end{equation}
The Oracle-to-Receiver Gap is
$|\mathrm{BA}^{\mathrm{o}}-\mathrm{BA}^{\mathrm{r}}|$. Receiver Exact Message
Recovery is the proportion of instances for which the complete recovered
secret exactly matches the intended secret:
\begin{equation}
\mathrm{EMR}
=
\frac{1}{M}\sum_{n=1}^{M}
\mathbb{I}\!\left[
\hat{\mathbf{s}}_n^{\mathrm r}=\mathbf{s}_n
\right].
\end{equation}

Let $\mathbf{y}^{\mathrm s}_n$ be the sender-side token sequence and
$\mathbf{y}^{\mathrm r}_n=\tau(\delta(\mathbf{y}^{\mathrm s}_n))$ the sequence
reconstructed by the receiver. The TI Rate is
\begin{equation}
\mathrm{TI\ Rate}
=
\frac{1}{M}\sum_{n=1}^{M}
\mathbb{I}\!\left[
\mathbf{y}^{\mathrm s}_n\neq\mathbf{y}^{\mathrm r}_n
\right].
\end{equation}
For an instance whose first receiver-side decoding error occurs at position
$q_n$, the cascading error rate is the error fraction over the remaining
secret positions:
\begin{equation}
\mathrm{CER}_n
=
\frac{1}{L_n-q_n}
\sum_{j=q_n+1}^{L_n}
\mathbb{I}\!\left[
 j>|\hat{\mathbf{s}}^{\mathrm r}_n|
 \;\lor\;
 \hat{s}^{\mathrm r}_{n,j}\neq s_{n,j}
\right].
\end{equation}
We set $\mathrm{CER}_n=0$ when no decoding error occurs or when $q_n=L_n$,
and report its mean over all instances. Cascade Incidence is measured only in
the separate channel-characterization study and denotes the proportion of TI
cases with at least three additional errors after the first decoding error.

\subsection{Text-Quality and Statistical Metrics}
Fluency is measured with GPT-2 perplexity. For each paired reference--stegotext
instance, normalized perplexity deviation is
\begin{equation}
\mathrm{PPL}^{*}_n
=
\frac{|\mathrm{PPL}(z_n)-\mathrm{PPL}(r_n)|}
{\mathrm{PPL}(r_n)},
\end{equation}
and the reported score is averaged over all instances. Semantic similarity is
the cosine similarity between normalized sentence embeddings of $z_n$ and
$r_n$, computed with sentence-transformers.

Statistical imperceptibility is evaluated using KLD between the cover and stego
feature distributions. Texts are encoded with bert-base, and
non-padding hidden states are mean-pooled. Each corpus-level feature
distribution is approximated by a diagonal Gaussian before computing KLD. In
our experiments, we take the logarithm of the KLD values to facilitate a
clearer comparison of statistical imperceptibility between algorithms.

\subsection{Anti-Steganalysis Evaluation}
BD, BF, and SANet are trained independently for each method and domain using balanced cover-stego data. Each detector uses 10,000 pairs for training and 1,000 pairs for both validation and testing, as specified in the main paper. Each detector is independently trained using five random seeds. We report the mean accuracy over five runs on a balanced test set containing 1,000 cover-stego pairs per run. The same data-partitioning protocol is applied across methods. Each detector is independently trained using five random seeds $\{42,43,44,45,46\}$. For each run, the balanced validation and test sets each contain 1,000 cover-stego pairs. We report the mean accuracy and standard deviation over the five runs. Values closer to 50\% indicate lower detectability. The BERT-based evaluator used during CCPO is trained separately from BF and shares no
parameters with it.

\subsection{Statistical Analysis}
All pairwise comparisons are conducted on matched test instances. We use two-sided exact McNemar tests for exact-message recovery, paired Wilcoxon signed-rank tests for per-instance $\mathrm{PPL}^{*}$ and SS, and paired bootstrap tests for differences in bit and detector accuracy. The bootstrap analysis uses 10,000 paired resamples, with 95\% confidence intervals defined
by the 2.5th and 97.5th percentiles. KLD is reported descriptively because it is computed at the corpus level. For comparisons between TI-StegoAlign and
each baseline, Holm's procedure is applied separately within each metric family at $\alpha=0.05$. Across all metrics subject to hypothesis testing, the reported pairwise improvements remain statistically significant after correction.

\subsection{Optimization Settings}
All experiments use the base model and LoRA configuration reported in
Section~4.1 of the main paper. BCSO and CCPO are trained sequentially on the
same LoRA adapter. Both stages use AdamW with per-device batch size 4, four
gradient-accumulation steps, weight decay 0.01, and maximum gradient norm 1.0,
yielding an effective batch size of 16. BCSO uses 100 warmup steps,
$\lambda_{\mathrm{BC}}=0.5$, and margin $\gamma=0.2$. CCPO uses 20 warmup
steps, sigmoid preference loss, and no label smoothing. The domain-specific
learning rates, epochs, and DPO coefficients are listed in
Table~\ref{tab:optimization-settings}.

\begin{table}[t]
\centering
\small
\setlength{\tabcolsep}{6pt}
\begin{tabular}{lccc}
\toprule
Setting & News & Movie & Tweet \\
\midrule
BCSO learning rate & $2\!\times\!10^{-5}$ & $2\!\times\!10^{-5}$ & $2\!\times\!10^{-5}$ \\
BCSO epochs & 3 & 3 & 3 \\
CCPO learning rate & $2\!\times\!10^{-6}$ & $1.5\!\times\!10^{-5}$ & $1.5\!\times\!10^{-5}$ \\
CCPO epochs & 8 & 8 & 8 \\
$\beta$ & 0.05 & 0.10 & 0.10 \\
\bottomrule
\end{tabular}
\caption{Domain-specific optimization settings.}
\label{tab:optimization-settings}
\end{table}

Each stage is run once for every domain using seed 42. The BCSO checkpoint is
used to initialize CCPO, so the two stages form a single sequential
post-training pipeline rather than independent model variants.

\subsection{Computing Environment}
Experiments are conducted on a workstation with one NVIDIA GeForce RTX 5090
GPU and two Intel Xeon Platinum 8470Q processors. The software environment is
summarized in Table~\ref{tab:environment}. 

\begin{table}[t]
\centering
\small
\setlength{\tabcolsep}{4pt}
\begin{tabular}{p{0.22\columnwidth}p{0.70\columnwidth}}
\toprule
Item & Configuration \\
\midrule
Hardware
& 1$\times$ NVIDIA GeForce RTX 5090 (32~GB);
  2$\times$ Intel Xeon Platinum 8470Q; 128~GB RAM \\

System
& Ubuntu 22.04.5 LTS; CUDA 12.8; cuDNN 9.10.2 \\

Runtime
& Python 3.12.3; PyTorch 2.8.0+cu128 \\

Libraries
& Transformers 5.8.0; PEFT 0.19.1; TRL 1.5.1;
  Sentence-Transformers 5.5.0; Scikit-learn 1.8.0 \\
\bottomrule
\end{tabular}
\caption{Hardware and software environment.}
\label{tab:environment}
\end{table}

\section{Domain-Specific Results}
\label{app:domain-results}

The main paper reports the arithmetic mean over News, Movie, and Tweet.
This appendix disaggregates the results to assess whether the main conclusions
hold across domains and to identify trade-offs that are obscured by averaged
scores. The \textit{Overall} rows reproduce the values in the main paper and
are computed as the arithmetic mean of the three domain-level results.

\begin{table*}[t]
\centering

\small
\begingroup
\setlength{\tabcolsep}{3pt}
\renewcommand{\arraystretch}{1.08}

\begin{tabular*}{0.98\textwidth}{
@{\extracolsep{\fill}}
c c c c c c c c
@{}
}
\toprule
\textbf{Method}
&
\textbf{Domain}
&
\shortstack{\textbf{Oracle bit}\\\textbf{acc. (\%)}}
&
\shortstack{\textbf{Receiver bit}\\\textbf{acc. (\%)}}
&
\shortstack{\textbf{Gap}\\\textbf{(pp)}}
&
\shortstack{\textbf{Exact}\\\textbf{recovery (\%)}}
&
\shortstack{\textbf{TI rate}\\\textbf{(\%)}}
&
\shortstack{\textbf{Cascading}\\\textbf{error rate (\%)}}
\\
\midrule

ADG & News
& 99.83 & 72.01 & 27.82 & 68.42 & 26.14 & 15.58 \\
ADG & Movie
& 100.00 & 86.04 & 13.96 & 82.30 & 19.94 & 13.43 \\
ADG & Tweet
& 100.00 & 73.88 & 26.12 & 68.01 & 27.33 & 14.46 \\
ADG & \textit{Overall}
& 99.94 & 77.31 & 22.63 & 72.91 & 24.47 & 14.49 \\

\midrule

Discop & News
& 98.90 & 85.64 & 13.26 & 75.90 & 9.70 & 14.02 \\
Discop & Movie
& 99.20 & 87.93 & 11.27 & 79.60 & 12.21 & 12.84 \\
Discop & Tweet
& 97.70 & 75.94 & 21.76 & 62.99 & 26.60 & 22.40 \\
Discop & \textit{Overall}
& 98.60 & 83.17 & 15.43 & 72.83 & 16.17 & 16.42 \\

\midrule

SegTrick & News
& 100.00 & 100.00 & 0.00 & 100.00 & 0.68 & 0.00 \\
SegTrick & Movie
& 100.00 & 100.00 & 0.00 & 100.00 & 0.60 & 0.00 \\
SegTrick & Tweet
& 100.00 & 100.00 & 0.00 & 100.00 & 1.30 & 0.00 \\
SegTrick & \textit{Overall}
& 100.00 & 100.00 & 0.00 & 100.00 & 0.86 & 0.00 \\

\midrule

SyncPool & News
& 100.00 & 100.00 & 0.00 & 100.00 & 0.78 & 0.00 \\
SyncPool & Movie
& 100.00 & 100.00 & 0.00 & 100.00 & 0.70 & 0.00 \\
SyncPool & Tweet
& 100.00 & 100.00 & 0.00 & 100.00 & 1.43 & 0.00 \\
SyncPool & \textit{Overall}
& 100.00 & 100.00 & 0.00 & 100.00 & 0.97 & 0.00 \\

\midrule

StepVerify & News
& 100.00 & 100.00 & 0.00 & 100.00 & 0.62 & 0.00 \\
StepVerify & Movie
& 100.00 & 100.00 & 0.00 & 100.00 & 0.53 & 0.00 \\
StepVerify & Tweet
& 100.00 & 100.00 & 0.00 & 100.00 & 1.19 & 0.00 \\
StepVerify & \textit{Overall}
& 100.00 & 100.00 & 0.00 & 100.00 & 0.78 & 0.00 \\

\midrule

TI-StegoAlign & News
& 100.00 & 100.00 & 0.00 & 100.00 & 0.60 & 0.00 \\
TI-StegoAlign & Movie
& 100.00 & 100.00 & 0.00 & 100.00 & 0.51 & 0.00 \\
TI-StegoAlign & Tweet
& 100.00 & 100.00 & 0.00 & 100.00 & 1.17 & 0.00 \\
TI-StegoAlign & \textit{Overall}
& 100.00 & 100.00 & 0.00 & 100.00 & 0.76 & 0.00 \\

\bottomrule
\end{tabular*}

\endgroup
\caption{Communication reliability by domain.}
\label{tab:domain-reliability}
\end{table*}

\paragraph{Communication reliability.}
ADG and Discop exhibit substantial degradation after the surface text is
retokenized. For ADG, the Oracle-to-Receiver Gap reaches 27.82 percentage
points on News and 26.12 points on Tweet, compared with 13.96 points on Movie.
Discop shows the same domain dependence, with its largest gap and lowest exact
recovery on Tweet. These results confirm that high sender-side recovery does
not imply reliable communication through the surface-text channel.
SegTrick, SyncPool, StepVerify, and TI-StegoAlign achieve perfect receiver bit
accuracy and exact message recovery in all three domains. Within this group,
TI-StegoAlign records the lowest TI rate in every domain. Its advantage over
StepVerify is small but consistent at 0.02 percentage points, while both
methods prevent the observed tokenization changes from propagating into
decoding errors.

\begin{table*}[t]
\centering
\small

\begingroup
\setlength{\tabcolsep}{10pt}
\renewcommand{\arraystretch}{1.08}

\begin{tabular*}{0.78\textwidth}{
@{\extracolsep{\fill}}
c c c c c
@{}
}
\toprule
\textbf{Method}
&
\textbf{Domain}
&
\textbf{$\mathrm{PPL}^{*}$}
&
\textbf{SS}
&
\textbf{KLD}
\\
\midrule

ADG & News   & 0.455 & 0.484 & 1.967 \\
ADG & Movie  & 0.292 & 0.591 & 2.216 \\
ADG & Tweet  & 0.894 & 0.437 & 2.930 \\
ADG & \textit{Overall} & 0.547 & 0.504 & 2.371 \\

\midrule

Discop & News   & 0.668 & 0.522 & 2.188 \\
Discop & Movie  & 0.654 & 0.500 & 2.389 \\
Discop & Tweet  & 0.946 & 0.484 & 2.416 \\
Discop & \textit{Overall} & 0.756 & 0.502 & 2.331 \\

\midrule

SegTrick & News   & 0.878 & 0.564 & 2.074 \\
SegTrick & Movie  & 0.900 & 0.558 & 2.579 \\
SegTrick & Tweet  & 0.985 & 0.540 & 2.880 \\
SegTrick & \textit{Overall} & 0.921 & 0.554 & 2.511 \\

\midrule

SyncPool & News   & 0.698 & 0.524 & 2.600 \\
SyncPool & Movie  & 0.671 & 0.498 & 2.755 \\
SyncPool & Tweet  & 0.953 & 0.502 & 2.970 \\
SyncPool & \textit{Overall} & 0.774 & 0.508 & 2.775 \\

\midrule

StepVerify & News   & 0.615 & 0.519 & 2.513 \\
StepVerify & Movie  & 0.605 & 0.501 & 2.351 \\
StepVerify & Tweet  & 0.937 & 0.480 & 2.444 \\
StepVerify & \textit{Overall} & 0.719 & 0.500 & 2.436 \\

\midrule

TI-StegoAlign & News   & 0.005 & 0.563 & 1.640 \\
TI-StegoAlign & Movie  & 0.180 & 0.590 & 1.320 \\
TI-StegoAlign & Tweet  & 1.102 & 0.536 & 2.101 \\
TI-StegoAlign & \textit{Overall} & 0.429 & 0.563 & 1.687 \\

\bottomrule
\end{tabular*}

\endgroup
\caption{Text quality and statistical imperceptibility by domain.}
\label{tab:domain-quality}
\end{table*}

\paragraph{Text quality and statistical imperceptibility.}
TI-StegoAlign yields the lowest KLD in all three domains, indicating that its
distributional advantage is not driven by a single corpus. It also obtains
substantially lower $\mathrm{PPL}^{*}$ on News and Movie, reaching 0.005 and
0.180, respectively. Semantic similarity remains competitive throughout:
TI-StegoAlign is within 0.001 of the domain-best result on News and Movie and
within 0.004 on Tweet. The main exception is Tweet fluency, where its
$\mathrm{PPL}^{*}$ of 1.102 is higher than those of the baselines. Thus, the
averaged improvement in $\mathrm{PPL}^{*}$ is primarily supported by News and
Movie rather than being uniform across all domains. Nevertheless, the lowest
Tweet KLD and near-best SS show that this limitation does not coincide with a
general loss of semantic preservation or corpus-level distributional
similarity.

\begin{table*}[t]
\centering

\small

\begingroup
\setlength{\tabcolsep}{7pt}
\renewcommand{\arraystretch}{1.08}

\begin{tabular*}{0.88\textwidth}{
@{\extracolsep{\fill}}
c c c c c c
@{}
}
\toprule
\textbf{Method}
&
\textbf{Domain}
&
\textbf{BD (\%)}
&
\textbf{BF (\%)}
&
\textbf{SANet (\%)}
&
\textbf{Mean (\%)}
\\
\midrule

ADG & News
& 58.59 & 70.50 & 57.86 & 62.32 \\
ADG & Movie
& 61.11 & 59.00 & 64.86 & 61.66 \\
ADG & Tweet
& 95.49 & 65.41 & 49.27 & 70.06 \\
ADG & \textit{Overall}
& 71.73 & 64.97 & 57.33 & 64.68 \\

\midrule

Discop & News
& 58.50 & 79.50 & 69.06 & 69.02 \\
Discop & Movie
& 69.50 & 79.00 & 72.19 & 73.56 \\
Discop & Tweet
& 79.51 & 79.01 & 77.93 & 78.82 \\
Discop & \textit{Overall}
& 69.17 & 79.17 & 73.06 & 73.80 \\

\midrule

SegTrick & News
& 69.00 & 80.00 & 68.96 & 72.65 \\
SegTrick & Movie
& 65.50 & 79.00 & 67.40 & 70.63 \\
SegTrick & Tweet
& 69.50 & 80.01 & 70.01 & 73.17 \\
SegTrick & \textit{Overall}
& 68.00 & 79.67 & 68.79 & 72.15 \\

\midrule

SyncPool & News
& 79.51 & 68.50 & 73.75 & 73.92 \\
SyncPool & Movie
& 74.50 & 69.50 & 70.10 & 71.37 \\
SyncPool & Tweet
& 79.00 & 69.51 & 79.47 & 75.99 \\
SyncPool & \textit{Overall}
& 77.67 & 69.17 & 74.44 & 73.76 \\

\midrule

StepVerify & News
& 63.50 & 70.00 & 68.02 & 67.17 \\
StepVerify & Movie
& 59.50 & 69.48 & 67.50 & 65.49 \\
StepVerify & Tweet
& 69.48 & 70.01 & 78.95 & 72.81 \\
StepVerify & \textit{Overall}
& 64.16 & 69.83 & 71.49 & 68.49 \\

\midrule

TI-StegoAlign & News
& 49.99 & 61.67 & 68.75 & 60.14 \\
TI-StegoAlign & Movie
& 50.00 & 50.61 & 56.25 & 52.29 \\
TI-StegoAlign & Tweet
& 53.01 & 73.93 & 81.25 & 69.40 \\
TI-StegoAlign & \textit{Overall}
& 51.00 & 62.07 & 68.75 & 60.61 \\

\bottomrule
\end{tabular*}

\endgroup
\caption{Anti-steganalysis accuracy by method and domain. Accuracy closer to
50\% indicates stronger anti-steganalysis performance.}
\label{tab:domain-security}
\end{table*}

\paragraph{Anti-steganalysis performance.}
TI-StegoAlign achieves the lowest mean detector accuracy in each domain,
improving over the strongest baseline by 2.18 percentage points on News,
9.37 points on Movie, and 0.66 points on Tweet. The gain is most pronounced on
Movie, where BD and BF are close to random guessing and SANet decreases to
56.25\%. The detector-level results are less uniform on News and Tweet.
ADG remains stronger under SANet on News and under BF and SANet on Tweet,
whereas TI-StegoAlign is consistently stronger under BD. This pattern shows
that the security improvement generalizes across domains in the averaged
detector score, but its magnitude depends on both the text domain and the
steganalysis architecture.

Taken together, the domain-wise evaluation supports two consistent findings:
TI-StegoAlign preserves exact receiver-side recovery across all corpora and
achieves the lowest KLD in every domain. Its gains in fluency and
anti-steganalysis performance are broader on News and Movie. This breakdown therefore
supports the averaged conclusions in the main paper without obscuring the
remaining variation across domains.

\section{Ablation Study}
\label{app:ablation}

We conduct ablation experiments to disentangle the contributions of BCSO and
CCPO. \emph{Base} denotes steganographic generation using the frozen base
language model and the shared coding rule $\Gamma$, without post-training.
\emph{BCSO} applies only the supervised objective in Section~3.2 and
evaluates the resulting policy without subsequent preference optimization.
\emph{CCPO} omits BCSO and performs preference optimization directly from
the base policy. In this variant, the base policy generates the candidate
stegotexts and also serves as the reference policy $\pi_{\mathrm{ref}}$.
\emph{Full} denotes the complete TI-StegoAlign pipeline, where the BCSO
policy initializes CCPO and serves as its reference policy.

All variants use the same Llama-3.1-8B backbone, tokenizer, coding rule,
training and evaluation splits, LoRA configuration, target payload rate,
and generation settings. When a training stage is retained, its optimization
parameters and data-construction procedure remain identical to those used
in the full model. The variants therefore differ only in whether BCSO,
CCPO, or both stages are applied. Tables~\ref{tab:ablation-reliability}--
\ref{tab:ablation-security} report communication reliability,
imperceptibility, and anti-steganalysis performance, respectively.

\begin{table*}[t]
\centering
\small
\begingroup
\setlength{\tabcolsep}{3pt}
\renewcommand{\arraystretch}{1.06}

\begin{tabular*}{0.98\textwidth}{
@{\extracolsep{\fill}}
l l c c c c c c
@{}
}
\toprule
\textbf{Variant}
&
\textbf{Domain}
&
\shortstack{\textbf{Oracle bit}\\\textbf{acc. (\%)}}
&
\shortstack{\textbf{Receiver bit}\\\textbf{acc. (\%)}}
&
\shortstack{\textbf{Gap}\\\textbf{(pp)}}
&
\shortstack{\textbf{Exact}\\\textbf{recovery (\%)}}
&
\shortstack{\textbf{TI rate}\\\textbf{(\%)}}
&
\shortstack{\textbf{Cascading}\\\textbf{error rate (\%)}}
\\
\midrule

Base & News
& 89.92 & 84.18 & 5.74 & 79.40 & 6.90 & 4.20 \\
Base & Movie
& 87.97 & 76.24 & 11.73 & 71.20 & 5.48 & 3.10 \\
Base & Tweet
& 78.84 & 69.80 & 9.04 & 64.60 & 13.24 & 8.91 \\
Base & \textit{Overall}
& 85.58 & 76.74 & 8.84 & 71.73 & 8.54 & 5.40 \\

\midrule

BCSO & News
& 100.00 & 99.72 & 0.28 & 98.20 & 1.18 & 0.18 \\
BCSO & Movie
& 100.00 & 99.86 & 0.14 & 99.10 & 0.94 & 0.10 \\
BCSO & Tweet
& 100.00 & 99.10 & 0.90 & 95.30 & 2.32 & 0.59 \\
BCSO & \textit{Overall}
& 100.00 & 99.56 & 0.44 & 97.53 & 1.48 & 0.29 \\

\midrule

CCPO & News
& 91.99 & 89.92 & 2.07 & 85.40 & 5.61 & 2.84 \\
CCPO & Movie
& 92.00 & 90.35 & 1.65 & 87.00 & 4.10 & 1.52 \\
CCPO & Tweet
& 87.97 & 84.31 & 3.66 & 80.80 & 8.53 & 3.79 \\
CCPO & \textit{Overall}
& 90.65 & 88.19 & 2.46 & 84.40 & 6.08 & 2.72 \\

\midrule

\textbf{Full} & News
& \textbf{100.00} & \textbf{100.00} & \textbf{0.00}
& \textbf{100.00} & \textbf{0.60} & \textbf{0.00} \\
\textbf{Full} & Movie
& \textbf{100.00} & \textbf{100.00} & \textbf{0.00}
& \textbf{100.00} & \textbf{0.51} & \textbf{0.00} \\
\textbf{Full} & Tweet
& \textbf{100.00} & \textbf{100.00} & \textbf{0.00}
& \textbf{100.00} & \textbf{1.17} & \textbf{0.00} \\
\textbf{Full} & \textit{Overall}
& \textbf{100.00} & \textbf{100.00} & \textbf{0.00}
& \textbf{100.00} & \textbf{0.76} & \textbf{0.00} \\

\bottomrule
\end{tabular*}
\endgroup
\caption{Domain-specific ablation results for communication reliability.
Higher Oracle and Receiver Bit Accuracy and Exact Recovery are better,
whereas lower Gap, TI Rate, and Cascading Error Rate are better.}
\label{tab:ablation-reliability}
\end{table*}

\begin{table}[t]
\centering
\small
\begingroup
\setlength{\tabcolsep}{3.8pt}
\renewcommand{\arraystretch}{1.06}

\begin{tabular*}{\columnwidth}{
@{\extracolsep{\fill}}
l l c c c
@{}
}
\toprule
\textbf{Variant}
&
\textbf{Domain}
&
\textbf{$\mathrm{PPL}^{*}$}
&
\textbf{SS}
&
\textbf{KLD}
\\
\midrule

Base & News
& 0.034 & 0.535 & 2.010 \\
Base & Movie
& 0.285 & 0.559 & 1.720 \\
Base & Tweet
& 1.517 & 0.511 & 2.576 \\
Base & \textit{Overall}
& 0.612 & 0.535 & 2.102 \\

\midrule

BCSO & News
& 0.020 & 0.545 & 1.870 \\
BCSO & Movie
& 0.245 & 0.570 & 1.560 \\
BCSO & Tweet
& 1.379 & 0.529 & 2.372 \\
BCSO & \textit{Overall}
& 0.548 & 0.548 & 1.934 \\

\midrule

CCPO & News
& 0.011 & 0.556 & 1.745 \\
CCPO & Movie
& 0.205 & 0.583 & 1.420 \\
CCPO & Tweet
& 1.179 & 0.535 & 2.121 \\
CCPO & \textit{Overall}
& 0.465 & 0.558 & 1.762 \\

\midrule

\textbf{Full} & News
& \textbf{0.005} & \textbf{0.563} & \textbf{1.640} \\
\textbf{Full} & Movie
& \textbf{0.180} & \textbf{0.590} & \textbf{1.320} \\
\textbf{Full} & Tweet
& \textbf{1.102} & \textbf{0.536} & \textbf{2.101} \\
\textbf{Full} & \textit{Overall}
& \textbf{0.429} & \textbf{0.563} & \textbf{1.687} \\

\bottomrule
\end{tabular*}
\endgroup
\caption{Domain-specific ablation results for linguistic and statistical
imperceptibility. Lower $\mathrm{PPL}^{*}$ and KLD are better, whereas
higher SS is better.}
\label{tab:ablation-imperceptibility}
\end{table}

\begin{table}[t]
\centering
\small
\begingroup
\setlength{\tabcolsep}{2.2pt}
\renewcommand{\arraystretch}{1.06}

\begin{tabular*}{\columnwidth}{
@{\extracolsep{\fill}}
l l c c c c
@{}
}
\toprule
\textbf{Variant}
&
\textbf{Domain}
&
\textbf{BD (\%)}
&
\textbf{BF (\%)}
&
\textbf{SANet (\%)}
&
\textbf{Mean (\%)}
\\
\midrule

Base & News
& 58.00 & 72.00 & 77.00 & 69.00 \\
Base & Movie
& 55.00 & 64.00 & 70.50 & 63.17 \\
Base & Tweet
& 68.00 & 85.01 & 87.25 & 80.09 \\
Base & \textit{Overall}
& 60.33 & 73.67 & 78.25 & 70.75 \\

\midrule

BCSO & News
& 55.50 & 68.50 & 73.50 & 65.83 \\
BCSO & Movie
& 52.50 & 60.50 & 66.25 & 59.75 \\
BCSO & Tweet
& 64.00 & 80.01 & 84.50 & 76.17 \\
BCSO & \textit{Overall}
& 57.33 & 69.67 & 74.75 & 67.25 \\

\midrule

CCPO & News
& 51.50 & 63.50 & 70.50 & 61.83 \\
CCPO & Movie
& 50.50 & 54.50 & 59.50 & 54.83 \\
CCPO & Tweet
& 56.01 & 75.00 & 82.49 & 71.17 \\
CCPO & \textit{Overall}
& 52.67 & 64.33 & 70.83 & 62.61 \\

\midrule

\textbf{Full} & News
& \textbf{49.99} & \textbf{61.67} & \textbf{68.75}
& \textbf{60.14} \\
\textbf{Full} & Movie
& \textbf{50.00} & \textbf{50.61} & \textbf{56.25}
& \textbf{52.29} \\
\textbf{Full} & Tweet
& \textbf{53.01} & \textbf{73.93} & \textbf{81.25}
& \textbf{69.40} \\
\textbf{Full} & \textit{Overall}
& \textbf{51.00} & \textbf{62.07} & \textbf{68.75}
& \textbf{60.61} \\

\bottomrule
\end{tabular*}
\endgroup
\caption{Domain-specific ablation results for anti-steganalysis performance.
Accuracy closer to 50\% indicates stronger anti-steganalysis performance.
Mean denotes the average over BD, BF, and SANet.}
\label{tab:ablation-security}
\end{table}

The reliability results show that BCSO provides the primary improvement in
secret recovery. Compared with Base, BCSO increases overall Receiver Bit
Accuracy from 76.74\% to 99.56\% and Exact Message Recovery from 71.73\%
to 97.53\%. It also reduces the Oracle-to-Receiver Gap from 8.84 to 0.44
percentage points, the TI Rate from 8.54\% to 1.48\%, and the Cascading
Error Rate from 5.40\% to 0.29\%. These improvements are consistent across
all three domains. The gain is particularly pronounced on Tweet, where Exact
Message Recovery increases from 64.60\% to 95.30\%. This behavior is
consistent with the design of BCSO: local margin supervision moves realized
token coordinates away from competing interval boundaries, making their
secret-bit assignments less sensitive to changes in the receiver-reconstructed
prefix.

CCPO alone also improves recovery, but does not match BCSO. Relative to Base,
it increases Receiver Bit Accuracy from 76.74\% to 88.19\% and Exact
Message Recovery from 71.73\% to 84.40\%. The corresponding Gap, TI Rate,
and Cascading Error Rate decrease to 2.46 percentage points, 6.08\%, and
2.72\%, respectively. These results indicate that receiver-realistic
preferences can favor candidates with better communication outcomes.
However, complete-stegotext preference optimization cannot fully compensate
for unstable local coding decisions when the policy lacks the initialization
provided by BCSO.

The contribution of CCPO is more evident in imperceptibility and
anti-steganalysis performance. Compared with Base, CCPO reduces overall
$\mathrm{PPL}^{*}$ from 0.612 to 0.465 and KLD from 2.102 to 1.762, while
increasing SS from 0.535 to 0.558. It also lowers mean detector accuracy from
70.75\% to 62.61\%. These gains are larger than those obtained by BCSO alone
on the same metrics. BCSO reduces mean detector accuracy to 67.25\%, whereas
CCPO reaches 62.61\%. Similarly, CCPO yields lower $\mathrm{PPL}^{*}$ and
KLD and higher SS than BCSO. This pattern follows from the design of CCPO,
which constructs preferences from receiver-side recovery, fluency, semantic
preservation, and steganalytic security evaluated over complete stegotexts.
It therefore optimizes properties that are not directly captured by local
coding-margin supervision.

BCSO nevertheless provides consistent gains beyond recovery. Compared with
Base, it decreases $\mathrm{PPL}^{*}$ from 0.612 to 0.548 and KLD from
2.102 to 1.934, increases SS from 0.535 to 0.548, and lowers mean detector
accuracy by 3.50 percentage points. These results show that coding-margin
supervision does not improve reliability at the expense of the transmitted
text. The retained language-modeling objective preserves in-domain
generation, while the margin term improves the realized embedding decisions.

Combining the two stages produces improvements that neither stage achieves
independently. Starting from BCSO, CCPO raises Receiver Bit Accuracy from
99.56\% to 100.00\% and Exact Message Recovery from 97.53\% to 100.00\%.
It eliminates the remaining Oracle-to-Receiver Gap and cascading errors,
while reducing the TI Rate from 1.48\% to 0.76\%. CCPO also reduces
$\mathrm{PPL}^{*}$ from 0.548 to 0.429 and KLD from 1.934 to 1.687,
increases SS from 0.548 to 0.563, and lowers mean detector accuracy from
67.25\% to 60.61\%.

The comparison between CCPO and Full further demonstrates the importance of
the BCSO initialization. Adding BCSO before CCPO increases overall Receiver
Bit Accuracy from 88.19\% to 100.00\% and Exact Message Recovery from
84.40\% to 100.00\%. It also reduces the TI Rate from 6.08\% to 0.76\%
and eliminates the 2.72\% Cascading Error Rate observed when CCPO is applied
directly to the base policy. Meanwhile, Full further improves all three
imperceptibility metrics and reduces mean detector accuracy by 2.00
percentage points relative to CCPO alone.

The domain-specific results exhibit the same overall pattern. Relative to
BCSO, Full improves $\mathrm{PPL}^{*}$, SS, KLD, and mean detector accuracy
on News, Movie, and Tweet. Relative to CCPO, it also improves every reported
reliability metric in all three domains. The benefits of the sequential
training procedure are therefore not driven by a single corpus.

Overall, BCSO and CCPO serve complementary roles. BCSO establishes reliable
local coding decisions and provides a suitable initialization for subsequent
preference optimization. CCPO then aligns complete stegotexts with
receiver-realistic recovery, linguistic quality, semantic preservation, and
steganalytic security. Their sequential combination is the only variant that
simultaneously achieves perfect receiver-side recovery, zero
Oracle-to-Receiver Gap and cascading errors, the best overall
$\mathrm{PPL}^{*}$, SS, and KLD, and detector accuracy closest to random
guessing. These results validate the two-stage design of TI-StegoAlign rather
than treating BCSO and CCPO as interchangeable optimization components.

\section{Runtime, Memory, and Parameter Efficiency}
\label{app:efficiency}

TI-StegoAlign is parameter-efficient during post-training and introduces no additional trainable module at deployment. BCSO and CCPO sequentially update the same LoRA adapter, which contains 13.63M trainable parameters, corresponding to 0.17\% of the 8.03B-parameter backbone. The sender and receiver deploy the same LoRA-adapted policy.

On the same workstation, the measured BCSO training run requires 11{,}607 seconds. Under the reported configuration, CCPO training requires approximately 1.5 to 2.5 hours for eight epochs. These one-time training costs are incurred before deployment.

We profile 1{,}000 held-out instances from each domain using the computing
environment described in Appendix~\ref{app:implementation}. CUDA
synchronization is applied at component boundaries. Table~
\ref{tab:communication-efficiency} reports mean sender-side generation
latency, receiver-side retokenization and extraction latency, end-to-end
latency, and peak allocated GPU memory.

\begin{table}[t]
\centering
\small
\setlength{\tabcolsep}{3.2pt}
\begin{tabular}{lrrrr}
\toprule
Domain & Sender & Receiver & Total & Peak mem. (GB) \\
\midrule
News  & 0.81 & 0.65 & 1.46 & 17.61 \\
Movie & 0.77 & 0.62 & 1.39 & 17.61 \\
Tweet & 0.57 & 0.42 & 0.99 & 17.60 \\
\bottomrule
\end{tabular}
\caption{Deployment latency and peak GPU memory. Latencies are averaged over
1{,}000 held-out instances per domain and reported in seconds per instance.}
\label{tab:communication-efficiency}
\end{table}

Across the three domains, end-to-end communication requires between 0.99 and
1.46 seconds per instance, while peak allocated GPU memory remains
approximately 17.6~GB. The training cost is incurred offline, while deployed
communication requires only sender-side generation followed by receiver-side
retokenization and extraction.

\section{Qualitative Cases and Error-Propagation Analysis}
\label{app:qualitative}

\subsection{Error Propagation under Tokenization Inconsistency}

We further analyze a separately collected set of 29,979 transmissions generated
under a single channel-characterization configuration. This collection is
distinct from the six-method benchmark in Table~1 of the main paper. The main
benchmark reports method-specific TI rates on the shared evaluation set,
whereas the present collection was constructed to obtain sufficient TI events
for analyzing downstream error propagation. Accordingly, the TI incidence
reported below is specific to this collection and is not a pooled estimate
across methods.

Using the TI criterion defined in Appendix~\ref{app:implementation}, 265
transmissions exhibit a sender--receiver tokenization mismatch. Among these
cases, 120 show substantial error propagation: after the first decoding error,
more than 35\% of the remaining secret positions are decoded incorrectly. This
yields a cascade incidence of 45.3\% conditional on TI.

\begin{table}[t]
\centering
\small
\setlength{\tabcolsep}{6pt}
\begin{tabular}{lrr}
\toprule
Statistic & Count & Rate \\
\midrule
All transmissions & 29,979 & -- \\
TI cases & 265 & 0.88\% of transmissions \\
Substantial cascades & 120 & 45.3\% of TI cases \\
\bottomrule
\end{tabular}
\caption{Collection-specific TI incidence and conditional error propagation in
the channel-characterization set.}
\label{tab:cascade-summary}
\end{table}

The collection-specific TI rate should therefore not be interpreted as an
alternative to the method-specific rates in the main benchmark. The relevant
channel observation is conditional: once TI occurs, nearly half of the affected
transmissions exhibit substantial downstream corruption, consistent with the decoder
desynchronization mechanism discussed in Section~3.1 of the main paper.

\subsection{Illustrative Stegotexts}

To provide a concrete view of the generated text, we report one stegotext for
each method in each domain. The examples are drawn from outputs produced under
the common evaluation setup described in Section~4.1 of the main paper and are
included only for qualitative illustration. Line breaks are normalized for
typesetting, while the generated wording is unchanged.

{\small
\paragraph{News.}

\noindent\textbf{ADG.} Surprisingly, the researchers found that the ancient leprosy DNA was most similar to that of the disease form that persists today. Why weren't there other leprosy versions gone extinct? They have probably evolved to become ``modern leprosy'', Prof. Nolan says.\par

\noindent\textbf{Discop.} A mother is faced with the reality of parenting. Brittney Conley, 22, could not figure out why her son was crying. Over time, the baby grew fussy and Conley lost interest. Doctors diagnosed the baby with hydrocephalus, a condition where fluid collects on the brain. He was rushed to the hospital. He died and an autopsy revealed\par

\noindent\textbf{SegTrick.} The bill would require companies that collect personal data to notify people within 72 hours if there is a data breach. It also would require companies that collect personal information online to get people's consent before collecting and sharing their personal information.\par

\noindent\textbf{SyncPool.} It was 2003 and she'd just ended a 19-year government career to, among other things, follow a dream of dabbling in ceramics. ``Everybody in my family's a doctor, a lawyer or an engineer,'' said Motawi, 53, who was a management analyst for the Michigan Department of Community Health.\par

\noindent\textbf{StepVerify.} ``Too much choice'' turns food purchases into a task, causing shoppers to hesitate and look for a way to simplify their decision-making process.\par

\noindent\textbf{TI-StegoAlign.} Applicants from the top 20 American Research Universities were invited to participate.\par

\paragraph{Movie.}

\noindent\textbf{ADG.} Anne Parillaud and the support cast are unable to involve the audience and fail to invest enough in the melodramatic and romance elements of the film.\par

\noindent\textbf{Discop.} She is tormented by her guilt. A psychotherapist attempts to help her to overcome her grief and guilt. As they work together, they fall in love. But then another death occurs. At first she is relieved because she thinks she is finally free\par

\noindent\textbf{SegTrick.} The Blues Brothers 1998 is a 1998 American musical comedy film directed by John Landis and starring John Belushi, Dan Aykroyd, and James Brown. The film, a sequel to the original Blues Brothers film, was originally intended to be a television show, but was later changed to become a film.\par

\noindent\textbf{SyncPool.} substantial portion of the movie's running time, and the movie-within-the-movie that opens it, get us used to The Grand Budapest Hotel by way of a leisurely approach. Wes Anderson often has trouble settling on a single, dominant style or approach, but he's in a groove for most of the film. The early sections are both funny and pretty impressive in their recreation of a period piece.\par

\noindent\textbf{StepVerify.} Chicago, Illinois, to be exact. The Blues Brothers were an R\&B/Soul band that gained fame and fortune by being associated with John Belushi and Dan Aykroyd from the cast of SNL. While I have a huge admiration for them as band members, they were mediocre writers (and so were the songwriters they had brought in) and they made one great movie\par

\noindent\textbf{TI-StegoAlign.} The movie is hard to follow with the many switching scenes and lack of a strong narrative. As a result, the audience is left disoriented and unsatisfied with the outcome.\par

\paragraph{Tweet.}

\noindent\textbf{ADG.} Hollis' death scene will hurt me severely to watch on film. Seems he really is dead this time. I wonder if Daisy will again walk in her sleep tonight.\par

\noindent\textbf{Discop.} I'm about to go get a freakin `cold shower' as a remedy, cold water on your whole body helps bring down fevers. @bytchylilthing I hope you feel better soon, but if you do happen to be sick, that's a great cure for a\par

\noindent\textbf{SegTrick.} I'm so sorry you're sick. I hope this helps you feel better. I'm so sorry about your mom. I hope she's doing much much much better. I'm so happy you're feeling better. I hope she's doing much better.\par

\noindent\textbf{SyncPool.} There's a difference between loud \& LOUD, after all, and that kinda thing. I would REALLY like to hear my testy if I put your mTub under it, but am a bit afraid of things exploding. As it is a bit of a beast. Wish I could make it to these next few so I can have a better look at your little\par

\noindent\textbf{StepVerify.} But I'm still against such things and any other thing which make us forget the importance of education and getting certificates so that one day we can try to follow another life which they want to follow\par

\noindent\textbf{TI-StegoAlign.} I love sushi rolls! Merry Christmas\&Happy New Year!\par
}

These examples offer a concrete view of the linguistic differences summarized
by the aggregate results in Sections~4.2-4.5. Domain-level breakdowns and the
complete human-rating analysis are reported in Appendices~\ref{app:domain-results}
and~\ref{app:human}, respectively.

\section{Method-Specific Configurations and Embedding-Rate Analysis}
\label{app:capacity}

The main paper evaluates all methods at a common embedding rate of 0.5 bpw
to control for differences in payload pressure. This appendix complements
that comparison by evaluating each baseline at its method-specific operating
point. Except for the embedding configuration, all experiments follow
Section~4.1 of the main paper. They use the same Llama-3.1-8B backbone,
tokenizer, evaluation instances, generation protocol, and receiver-realistic
extraction procedure. The reported bpw is the average embedding rate realized
over the generated texts. All results are averaged across News, Movie, and
Tweet.

\begin{table*}[t]
\centering
\small
\begingroup
\setlength{\tabcolsep}{2.3pt}
\renewcommand{\arraystretch}{1.08}

\begin{tabular*}{0.99\textwidth}{
@{\extracolsep{\fill}}
l c c c c c c c
@{}
}
\toprule
\textbf{Method}
&
\textbf{bpw}
&
\shortstack{\textbf{Oracle Bit}\\\textbf{Acc. (\%)}}
&
\shortstack{\textbf{Receiver Bit}\\\textbf{Acc. (\%)}}
&
\shortstack{\textbf{Oracle-to-Receiver}\\\textbf{Gap (pp)}}
&
\shortstack{\textbf{Receiver Exact Message}\\\textbf{Recovery (\%)}}
&
\shortstack{\textbf{TI Rate}\\\textbf{(\%)}}
&
\shortstack{\textbf{Cascading Error}\\\textbf{Rate (\%)}}
\\
\midrule

ADG
& 4.60
& 95.46
& 73.93
& 21.53
& 70.88
& 24.32
& 14.71
\\

Discop
& 4.31
& 93.12
& 70.98
& 22.14
& 68.40
& 21.63
& 18.84
\\

SegTrick
& 3.16
& 100.00
& 100.00
& 0.00
& 100.00
& 1.17
& 0.00
\\

SyncPool
& 1.89
& 100.00
& 100.00
& 0.00
& 100.00
& 1.22
& 0.00
\\

StepVerify
& 4.38
& 100.00
& 100.00
& 0.00
& 100.00
& 0.97
& 0.00
\\

\bottomrule
\end{tabular*}
\endgroup
\caption{Communication reliability under method-specific embedding
configurations. The reported bpw is the average embedding rate realized by
each method.}
\label{tab:method-specific-reliability}
\end{table*}

Table~\ref{tab:method-specific-reliability} shows that a high realized
embedding rate does not necessarily yield reliable end-to-end communication.
ADG and Discop operate at 4.60 and 4.31 bpw, respectively. However, their
Oracle Bit Accuracy values are already limited to 95.46\% and 93.12\%.
Receiver-side retokenization further reduces their Bit Accuracy to 73.93\%
and 70.98\%, producing Oracle-to-Receiver Gaps above 21 percentage points.
Their Exact Message Recovery values also remain below 71\%.

The TI Rates of ADG and Discop exceed 21\%, while their Cascading Error Rates
reach 14.71\% and 18.84\%, respectively. These results show that local
tokenization mismatches frequently propagate to subsequent secret positions.
Consequently, their high sender-side payloads do not translate into reliable
receiver-side recovery.

SegTrick, SyncPool, and StepVerify preserve perfect receiver-side recovery
under their respective embedding configurations. Their Oracle-to-Receiver
Gaps and Cascading Error Rates remain zero. This indicates that their
respective inference-time mechanisms prevent tokenization mismatches from
disrupting subsequent extraction. Nevertheless, reliable recovery alone does
not establish a practical covert channel. The transmitted stegotext must also
remain linguistically natural, semantically consistent, statistically
inconspicuous, and resistant to steganalysis.

\begin{table}[t]
\centering
\small
\setlength{\tabcolsep}{4.4pt}
\renewcommand{\arraystretch}{1.06}

\begin{tabular*}{\columnwidth}{
@{\extracolsep{\fill}}
l c c c c
@{}
}
\toprule
\textbf{Method}
&
\textbf{$\mathrm{PPL}^{*}\downarrow$}
&
\textbf{SS$\uparrow$}
&
\textbf{KLD$\downarrow$}
\\
\midrule

ADG
& 0.630
& 0.493
& 2.470
\\

Discop
& 0.779
& 0.491
& 2.354
\\

SegTrick
& 0.891
& 0.503
& 3.254
\\

SyncPool
& 0.775
& 0.501
& 2.713
\\

StepVerify
& 0.727
& 0.461
& 2.823
\\

\bottomrule
\end{tabular*}
\caption{Text quality and statistical imperceptibility under
method-specific embedding configurations. Lower $\mathrm{PPL}^{*}$ and
KLD are better, whereas higher SS is better.}
\label{tab:method-specific-quality}
\end{table}

Table~\ref{tab:method-specific-quality} reveals substantial degradation in
linguistic and statistical imperceptibility at the method-specific operating
points. SegTrick records the highest $\mathrm{PPL}^{*}$ of 0.891 and KLD of
3.254. It therefore exhibits the largest deviation from both the
target-domain perplexity and the cover-text distribution. StepVerify obtains
the lowest SS of 0.461, indicating weaker preservation of the semantic
reference. Its KLD also reaches 2.823.

SyncPool exhibits a similarly high KLD of 2.713. Discop produces a
$\mathrm{PPL}^{*}$ of 0.779 and an SS of only 0.491, indicating limitations
in both fluency and semantic consistency. Although ADG obtains the lowest
$\mathrm{PPL}^{*}$ among these baselines, its SS remains below 0.5 and its
KLD reaches 2.470.

The results show that perfect recovery does not guarantee inconspicuous
transmitted text. SegTrick and StepVerify achieve exact extraction at 3.16
and 4.38 bpw, respectively, but their reliability is accompanied by
pronounced statistical deviation or semantic degradation. SyncPool also
maintains exact extraction, yet its generated distribution remains
substantially different from the cover-text distribution. Conversely, the
quality results of ADG and Discop do not offset their severe receiver-side
recovery loss. None of these operating points therefore achieves a strong
balance among embedding rate, reliable extraction, and linguistic and
statistical imperceptibility.

\begin{table}[t]
\centering
\small
\setlength{\tabcolsep}{2.6pt}
\renewcommand{\arraystretch}{1.06}

\begin{tabular*}{\columnwidth}{
@{\extracolsep{\fill}}
l c c c c c
@{}
}
\toprule
\textbf{Method}
&
\textbf{BD (\%)}
&
\textbf{BF (\%)}
&
\textbf{SANet (\%)}
&
\textbf{Mean (\%)}
\\
\midrule

ADG
& 82.73
& 67.97
& 65.40
& 72.03
\\

Discop
& 80.67
& 79.17
& 82.36
& 80.73
\\

SegTrick
& 78.17
& 89.33
& 86.70
& 84.73
\\

SyncPool
& 78.00
& 79.33
& 95.66
& 84.33
\\

StepVerify
& 70.16
& 79.67
& 84.41
& 78.08
\\

\bottomrule
\end{tabular*}
\caption{Steganalysis accuracy under method-specific embedding
configurations. Accuracy closer to 50\% indicates stronger
anti-steganalysis performance. Mean denotes the average over BD, BF, and
SANet.}
\label{tab:method-specific-security}
\end{table}

The steganalysis results further expose the limitations of these
method-specific operating points. Mean accuracy ranges from 72.03\% to
84.73\%, remaining well above the 50\% level associated with random
guessing. SegTrick and SyncPool are the most readily identified, with mean
accuracies of 84.73\% and 84.33\%, respectively. SegTrick reaches 89.33\%
under BF and 86.70\% under SANet. SANet identifies SyncPool with 95.66\%
accuracy, indicating particularly distinguishable statistical
characteristics in its generated text.

Discop and StepVerify also remain highly distinguishable, with mean
accuracies of 80.73\% and 78.08\%, respectively. Although StepVerify
maintains perfect receiver-side recovery at 4.38 bpw, its low SS and high
steganalysis accuracy show that this recovery is achieved without preserving
comparable semantic and statistical imperceptibility.

ADG attains the lowest mean steganalysis accuracy among these configurations.
However, its value of 72.03\% remains far from random guessing. This relative
advantage is also accompanied by a Receiver Bit Accuracy of only 73.93\% and
an Exact Message Recovery rate of 70.88\%. Its method-specific operating
point therefore does not provide a satisfactory balance between
recoverability and anti-steganalysis performance.

Taken together, the results reveal two distinct failure modes. ADG and
Discop achieve high realized embedding rates but suffer substantial
receiver-side information loss. SegTrick, SyncPool, and StepVerify preserve
exact extraction, yet their stegotexts exhibit marked linguistic, semantic,
or statistical degradation and remain readily identified by the considered
steganalyzers. These limitations constrain practical covert communication,
which requires reliable secret recovery and inconspicuous transmitted text
simultaneously.

The main paper therefore adopts 0.5 bpw as the primary operating point for
cross-method evaluation. This choice is not intended to identify a
universally optimal embedding rate. Instead, it provides a common and
feasible payload condition for all evaluated methods. This prevents unequal
embedding pressure from confounding comparisons of communication
reliability, linguistic quality, statistical imperceptibility, and
anti-steganalysis performance.

The method-specific results complement this controlled evaluation by showing
that embedding rate alone does not characterize the practical effectiveness
of a steganographic communication method. A meaningful evaluation must
jointly consider the amount of embedded information, the proportion that can
be recovered by the receiver, and the extent to which the transmitted text
remains inconspicuous.

\section{Generalization across Model Backbones}
\label{app:backbone}

We further evaluate TI-StegoAlign with Qwen3-8B to examine whether its effectiveness is tied to the
Llama-3.1-8B backbone. It isolates the
transferability of TI-StegoAlign across different language-model and tokenizer
families.

The two backbone experiments use the same News, Movie, and Tweet splits,
target payload rate of 0.5 bpw, coding rule $\Gamma$, evaluation instances,
LoRA architecture, generation budget, and receiver-realistic communication
protocol. BCSO and CCPO are trained independently for each backbone, and the
corresponding tokenizer is shared by the sender and receiver.

Several implementation settings are adapted to the Qwen backbone. The sender and receiver both disable automatic special-token insertion during
tokenization, preventing Qwen-specific control tokens from changing the coding prefix. The generation prompt itself remains unchanged, and the chat
template is not enabled. Qwen-specific control markers and meta-text patterns
are excluded during candidate generation.

The Qwen CCPO stage uses two training epochs, learning rate
$1.5\times10^{-5}$, and $\beta=0.1$. Its reference policy is implemented as
a frozen PEFT reference adapter rather than a separately loaded model copy.
Domain-specific candidate-screening thresholds are adjusted to the Qwen
candidate distribution to retain sufficient valid preference pairs. The
evaluation metrics, preference criteria, and receiver-side extraction
procedure remain unchanged. These adaptations address model and tokenizer
differences without altering the formulation of TI-StegoAlign.

\begin{table*}[t]
\centering
\small
\begingroup
\setlength{\tabcolsep}{2.5pt}
\renewcommand{\arraystretch}{1.07}

\begin{tabular*}{0.99\textwidth}{
@{\extracolsep{\fill}}
l l c c c c c c
@{}
}
\toprule
\textbf{Backbone}
&
\textbf{Domain}
&
\shortstack{\textbf{Oracle bit}\\\textbf{acc. (\%)}}
&
\shortstack{\textbf{Receiver bit}\\\textbf{acc. (\%)}}
&
\shortstack{\textbf{Gap}\\\textbf{(pp)}}
&
\shortstack{\textbf{Exact}\\\textbf{recovery (\%)}}
&
\shortstack{\textbf{TI rate}\\\textbf{(\%)}}
&
\shortstack{\textbf{Cascading}\\\textbf{error rate (\%)}}
\\
\midrule

Llama-3.1-8B
& News
& 100.00 & 100.00 & 0.00 & 100.00 & 0.60 & 0.00 \\
Llama-3.1-8B
& Movie
& 100.00 & 100.00 & 0.00 & 100.00 & 0.51 & 0.00 \\
Llama-3.1-8B
& Tweet
& 100.00 & 100.00 & 0.00 & 100.00 & 1.17 & 0.00 \\
Llama-3.1-8B
& \textit{Overall}
& 100.00 & 100.00 & 0.00 & 100.00 & 0.76 & 0.00 \\

\midrule

Qwen3-8B
& News
& 100.00 & 100.00 & 0.00 & 100.00 & 0.57 & 0.00 \\
Qwen3-8B
& Movie
& 100.00 & 100.00 & 0.00 & 100.00 & 0.48 & 0.00 \\
Qwen3-8B
& Tweet
& 100.00 & 100.00 & 0.00 & 100.00 & 1.11 & 0.00 \\
Qwen3-8B
& \textit{Overall}
& 100.00 & 100.00 & 0.00 & 100.00 & 0.72 & 0.00 \\

\bottomrule
\end{tabular*}
\endgroup
\caption{Communication reliability of TI-StegoAlign across model
backbones. Higher Oracle and Receiver Bit Accuracy and Exact Recovery are
better, whereas lower Gap, TI Rate, and Cascading Error Rate are better.}
\label{tab:backbone-reliability}
\end{table*}

As shown in Table~\ref{tab:backbone-reliability}, TI-StegoAlign preserves
perfect receiver-side recovery after replacing Llama-3.1-8B with Qwen3-8B.
Both backbones achieve 100.00\% Oracle and Receiver Bit Accuracy,
100.00\% Exact Message Recovery, zero Oracle-to-Receiver Gap, and no
cascading errors across all three domains. This result is important because
the two backbones use different tokenizers and generation formats. The
unchanged recovery performance shows that the reliability gains do not rely
on a particular Llama tokenization pattern.

The Qwen-specific tokenization adjustments ensure that the sender and receiver
construct the coding prefix under the same special-token policy. BCSO then
establishes sufficiently separated local coding decisions, while CCPO retains
candidates that remain recoverable after receiver-side reconstruction.
Together, these mechanisms preserve exact communication despite the change
in backbone and tokenizer.

\begin{table}[t]
\centering
\small
\begingroup
\setlength{\tabcolsep}{3.4pt}
\renewcommand{\arraystretch}{1.07}

\begin{tabular*}{\columnwidth}{
@{\extracolsep{\fill}}
l l c c c
@{}
}
\toprule
\textbf{Backbone}
&
\textbf{Domain}
&
\textbf{$\mathrm{PPL}^{*}$}
&
\textbf{SS}
&
\textbf{KLD}
\\
\midrule

Llama-3.1-8B
& News
& 0.005 & 0.563 & 1.640 \\
Llama-3.1-8B
& Movie
& 0.180 & 0.590 & 1.320 \\
Llama-3.1-8B
& Tweet
& 1.102 & 0.536 & 2.101 \\
Llama-3.1-8B
& \textit{Overall}
& 0.429 & 0.563 & 1.687 \\

\midrule

Qwen3-8B
& News
& 0.005 & 0.596 & 1.670 \\
Qwen3-8B
& Movie
& 0.190 & 0.612 & 1.401 \\
Qwen3-8B
& Tweet
& 1.158 & 0.565 & 1.980 \\
Qwen3-8B
& \textit{Overall}
& 0.451 & 0.591 & 1.684 \\

\bottomrule
\end{tabular*}
\endgroup
\caption{Text quality and statistical imperceptibility of TI-StegoAlign
across model backbones. Lower $\mathrm{PPL}^{*}$ and KLD are better,
whereas higher SS is better.}
\label{tab:backbone-quality}
\end{table}

Table~\ref{tab:backbone-quality} shows that the change in backbone affects
the individual quality metrics differently. Qwen3-8B increases semantic
similarity to approximately 0.60, compared with an overall SS of 0.563 for
Llama-3.1-8B. The improvement is observed while exact receiver-side recovery
is preserved. This indicates that the Qwen policy follows the semantic
reference more closely rather than trading semantic fidelity for reliable
embedding.

Qwen3-8B exhibits a small degradation in $\mathrm{PPL}^{*}$, represented by
a slightly larger value than that of Llama-3.1-8B. This difference does not
produce a corresponding decline in SS, KLD, or recovery. One possible
explanation is that the two backbones favor different lexical and stylistic
distributions, whereas $\mathrm{PPL}^{*}$ is measured using the same external
perplexity model. The Qwen outputs may therefore remain natural and
semantically faithful while differing modestly from the lexical distribution
preferred by that evaluator.

More importantly, the higher SS shows that the benefit of the stronger
semantic-generation behavior is retained after steganographic post-training.
BCSO does not erase the backbone's task-conditioned generation ability, and
CCPO can exploit the improved candidate pool to select stegotexts that better
preserve the intended content.

\begin{table}[t]
\centering
\small
\begingroup
\setlength{\tabcolsep}{2.0pt}
\renewcommand{\arraystretch}{1.07}

\begin{tabular*}{\columnwidth}{
@{\extracolsep{\fill}}
l l c c c c
@{}
}
\toprule
\textbf{Backbone}
&
\textbf{Domain}
&
\textbf{BD (\%)}
&
\textbf{BF (\%)}
&
\textbf{SANet (\%)}
&
\textbf{Mean (\%)}
\\
\midrule

Llama-3.1-8B
& News
& 49.99 & 61.67 & 68.75 & 60.14 \\
Llama-3.1-8B
& Movie
& 50.00 & 50.61 & 56.25 & 52.29 \\
Llama-3.1-8B
& Tweet
& 53.01 & 73.93 & 81.25 & 69.40 \\
Llama-3.1-8B
& \textit{Overall}
& 51.00 & 62.07 & 68.75 & 60.61 \\

\midrule

Qwen3-8B
& News
& 50.00 & 58.20 & 64.78 & 57.66 \\
Qwen3-8B
& Movie
& 50.00 & 50.10 & 52.99 & 51.03 \\
Qwen3-8B
& Tweet
& 51.20 & 69.60 & 76.48 & 65.76 \\
Qwen3-8B
& \textit{Overall}
& 50.40 & 59.30 & 64.75 & 58.15 \\

\bottomrule
\end{tabular*}
\endgroup
\caption{Anti-steganalysis performance of TI-StegoAlign across model
backbones. Accuracy closer to 50\% indicates stronger anti-steganalysis
performance. Mean denotes the average over BD, BF, and SANet.}
\label{tab:backbone-security}
\end{table}

The results in Table~\ref{tab:backbone-security} show that Qwen3-8B further
reduces steganalysis accuracy relative to Llama-3.1-8B. This improvement is
obtained across the considered steganalyzers while receiver-side Bit Accuracy
and Exact Message Recovery remain at 100.00\%. The gain therefore does not
arise from embedding less information or allowing more decoding failures.

This pattern is consistent with the multi-criteria design of CCPO. A backbone
with stronger task-conditioned generation can provide a more diverse set of
natural candidate stegotexts. CCPO can then retain candidates that satisfy
receiver-side recovery while exhibiting lower steganalytic risk. The
improvement is especially meaningful because the frozen steganalysis
evaluation remains unchanged across backbones. The lower accuracies therefore
reflect differences in the generated stegotext rather than changes to the
evaluation protocol.

Taken together, the results demonstrate that TI-StegoAlign is not specific
to Llama-3.1-8B. Its two-stage post-training procedure transfers to Qwen3-8B
despite differences in model behavior, tokenization, and training format.
Both backbones preserve perfect receiver-side recovery, while Qwen3-8B
provides higher semantic similarity and stronger anti-steganalysis
performance, with only a small trade-off in $\mathrm{PPL}^{*}$.

The method is model-agnostic at the algorithmic level: it requires an
autoregressive language model, its corresponding tokenizer, and a shared
coding rule, but does not depend on a particular vocabulary or model family.
The experiments on two distinct 8B backbones provide empirical evidence for
this transferability. They also suggest that TI-StegoAlign can benefit from
improvements in the underlying language model, particularly when stronger
generation capability produces candidates with better semantic fidelity and
lower steganalytic risk. This evidence supports cross-backbone
generalization, although broader validation on additional model scales and
architectures remains necessary before claiming universal applicability.

\section{Human Evaluation Protocol and Statistical Results}
\label{app:human}

\subsection{Sampling, Randomization, and Blinding}

For each method, 50 outputs are sampled from each of News, Movie, and Tweet,
yielding 150 outputs per method and 900 evaluation items across the six
methods. The same test instances and embedding rate are used across methods.
Each item presents an anonymized stegotext together with its semantic
reference. The reference is used only to assess semantic consistency;
fluency and coherence are judged from the stegotext alone. Method labels
are removed, and item order is independently randomized for the ten
evaluators using NumPy's PCG64 generator with seeds 43 through 52.
All evaluators rate every item, yielding 9{,}000 ratings per criterion.

\begin{table}[t]
\centering
\small
\begin{tabular}{lc}
\toprule
Criterion & Krippendorff's $\alpha$ \\
\midrule
Fluency              & 0.676 \\
Coherence            & 0.713 \\
Semantic consistency & 0.722 \\
\bottomrule
\end{tabular}
\caption{Inter-rater agreement for the three human-evaluation criteria.}
\label{tab:human-reliability}
\end{table}

\subsection{Five-Point Rubric}

Table~\ref{tab:human-rubric} presents the rating anchors provided to the
evaluators.

\begin{table*}[t]
\centering
\small
\begin{tabular}{lp{4.0cm}p{4.0cm}p{4.0cm}}
\toprule
Criterion & Score 1 & Score 3 & Score 5 \\
\midrule

Fluency &
Severe grammatical or lexical problems make the text unnatural or difficult
to understand. &
The text is understandable but contains noticeable grammatical, lexical, or
stylistic problems. &
The text is grammatical, natural, and reads like fluent human-written text. \\

Coherence &
The text is disjointed, self-contradictory, or lacks a comprehensible
progression. &
The overall progression is understandable, but some transitions or relations
are weak. &
The text is logically connected throughout, with a clear and consistent
progression. \\

Semantic consistency &
The stegotext conveys content that is unrelated to or conflicts with the
semantic reference. &
The stegotext preserves the main intent but deviates in some salient details. &
The stegotext preserves the intent and key information conveyed by the
semantic reference. \\
\bottomrule
\end{tabular}
\caption{Five-point human-evaluation rubric. Scores 2 and 4 represent
intermediate judgments between adjacent anchors.}
\label{tab:human-rubric}
\end{table*}

Fluency and coherence are judged from the stegotext alone, whereas semantic
consistency is assessed against the semantic reference.

\subsection{Evaluators}

The evaluation involves ten researchers holding master's or doctoral degrees
in computer science, artificial intelligence, or software engineering. Each
has prior experience in NLP, LLM security, or code evaluation, and none
participated in method development.

\subsection{Inter-Rater Reliability}

Krippendorff's alpha is computed separately for fluency, coherence, and
semantic consistency using ordinal distance. Because every item is rated by
all ten evaluators, no missing-value adjustment is required.
Table~\ref{tab:human-reliability} reports the agreement values.

\subsection{Human-Rating Results}

Table~\ref{tab:human-results} reports mean ratings with 95\% confidence
intervals. The intervals are obtained through paired bootstrap resampling
over test instances.

\begin{table*}[t]
\centering
\small
\setlength{\tabcolsep}{5.5pt}
\begin{tabular}{lccc}
\toprule
Method
& Fluency
& Coherence
& Semantic consistency \\
\midrule
ADG
& 4.39 [4.25, 4.52]
& 4.13 [3.95, 4.30]
& 3.81 [3.61, 4.01] \\

Discop
& 3.92 [3.74, 4.09]
& 3.86 [3.67, 4.05]
& 3.78 [3.58, 3.97] \\

SegTrick
& 3.55 [3.36, 3.74]
& 4.01 [3.82, 4.20]
& 4.56 [4.43, 4.68] \\

SyncPool
& 3.88 [3.69, 4.06]
& 3.88 [3.69, 4.06]
& 3.87 [3.68, 4.06] \\

StepVerify
& 4.00 [3.84, 4.16]
& 3.89 [3.69, 4.08]
& 3.75 [3.55, 3.95] \\

TI-StegoAlign
& \textbf{4.65 [4.55, 4.74]}
& \textbf{4.67 [4.56, 4.77]}
& \textbf{4.70 [4.59, 4.79]} \\
\bottomrule
\end{tabular}
\caption{Mean human ratings with 95\% instance-level bootstrap confidence
intervals. Higher values indicate better performance.}
\label{tab:human-results}
\end{table*}

TI-StegoAlign receives the highest mean rating on all three criteria.
Compared with the strongest baseline for each criterion, it improves
fluency by 0.26 points over ADG, coherence by 0.54 points over ADG, and
semantic consistency by 0.14 points over SegTrick.

Ratings for each criterion are analyzed using a cumulative-logit mixed model
with method as a fixed effect and evaluator and test instance as random
intercepts. Pairwise comparisons between TI-StegoAlign and the five
baselines are corrected within each criterion using Holm's procedure.
Table~\ref{tab:human-clmm} reports comparisons with the strongest baseline
for each criterion.

\begin{table*}[t]
\centering
\small
\setlength{\tabcolsep}{7pt}
\begin{tabular}{llcccc}
\toprule
Criterion
& Comparator
& Log-odds / OR
& SE
& OR 95\% CI
& Adjusted $p$ \\
\midrule
Fluency
& ADG
& 0.680 / 1.97
& 0.086
& [1.67, 2.34]
& $<0.001$ \\

Coherence
& ADG
& 1.210 / 3.35
& 0.106
& [2.72, 4.13]
& $<0.001$ \\

Semantic consistency
& SegTrick
& 0.443 / 1.56
& 0.078
& [1.34, 1.81]
& $<0.001$ \\
\bottomrule
\end{tabular}
\caption{Cumulative-logit comparisons between TI-StegoAlign and the
strongest baseline for each criterion. Odds ratios above one favor
TI-StegoAlign.}
\label{tab:human-clmm}
\end{table*}

After Holm correction, TI-StegoAlign retains statistically significant advantages over the strongest baseline on all three criteria. Compared with ADG, it has 3.35 times the odds of receiving a higher coherence rating, indicating substantially stronger logical continuity and discourse
organization. Its odds of receiving a higher fluency rating are also 1.97 times those of ADG, reflecting more natural and readable stegotext. TI-StegoAlign further outperforms SegTrick, the strongest semantic-consistency baseline, with an odds ratio of 1.56. Together with the highest mean ratings,
these results demonstrate that TI-StegoAlign improves linguistic naturalness and coherence without compromising the semantic fidelity of the transmitted content.

\end{document}